\documentclass[12pt,preprint]{aastex}
\usepackage{emulateapj5}
\usepackage{graphicx,natbib}

\begin{document} 

\title{{Halos of Spiral Galaxies. III. Metallicity Distributions\altaffilmark{1}}}

\author{M. Mouhcine\altaffilmark{2,3}, R.M. Rich\altaffilmark{2}, 
        H.C. Ferguson\altaffilmark{4}, T.M. Brown\altaffilmark{4}, 
        T.E. Smith\altaffilmark{4}}
\altaffiltext{1}{Based on observations with the NASA/ESA Hubble Space 
                 Telescope, obtained at the Space Telescope Science Institute,
		 which is operated by the Association of Universities
		 for Research in Astronomy, Inc.,under NASA contract 
		 NAS 5-26555}	
\altaffiltext{2}{Department of Physics and Astronomy, UCLA, Math-Science 
                 Building, 8979, Los Angeles, CA 90095-1562}
\altaffiltext{3}{Present address: School of Physics and Astronomy, 
                 University of Nottingham, 
                 University Park,  Nottingham NG7 2RD, UK}		 
\altaffiltext{4}{Space Telescope Science Institute, 3700, San Martin Drive,
                 Baltimore, MD, 21218, USA}

\begin{abstract}

We report results of a campaign to image the stellar populations
in the halos of highly inclined spiral galaxies, with the fields
roughly 10 kpc (projected) from the nuclei.  We use the F814W (I)
and F606W (V) filters in the Wide Field Planetary Camera 2,
on board the Hubble Space telescope.

We unambiguously resolve the stellar halos one to two magnitudes fainter
than the tip of the red giant branch. Extended halo populations are
detected in all galaxies. The color-magnitude diagrams appear to
be completely dominated by giant-branch stars, with no evidence
for the presence of young stellar populations in any of the fields.

The metallicity distribution function for the galaxy sample is derived 
from interpolation within an extensive grid of red giant branch loci.
These loci are derived from theoretical sequences which are calibrated 
using the Galactic globular clusters, and also using empirical sequences 
for metal-rich stellar populations. We find that the metallicity 
distribution functions are dominated by metal-rich populations, with 
a tail extending toward the metal poor end. To first order, the 
overall shapes of the metallicity distribution functions are similar 
to what is predicted by simple, single-component model of chemical 
evolution with the effective yields increasing with galaxy luminosity. 
However, metallicity distributions significantly narrower than the 
simple model are observed for a few of the most luminous galaxies in 
the sample. The discrepancies are similar to those previously observed 
for NGC~5128, the halo of M~31, and the Galactic bulge.

Our observations may be used to help distinguish between models for
the formation of spiral galaxies. It appears clear that more luminous 
spiral galaxies also have more metal-rich stellar halos.
The increasingly significant departures from the closed-box model 
for the more luminous galaxies indicate that a parameter in addition
to a single yield is required to describe chemical evolution. 
This parameter, which could be related to gas infall or outflow either 
in situ or in progenitor dwarf galaxies that later merge to form the 
stellar halo, tends to act to make the metallicity distributions 
narrower at high metallicity.

\end{abstract}


\section{ Introduction}

The oldest stellar populations in galaxies offer clues to the earliest
epochs of galaxy formation. The nature of halo stellar populations, detailed 
structures, and kinematics are clues to the understanding of how galaxies 
have assembled their mass. Much of what we know about such early populations 
comes from detailed studies of the stellar halo of the Milky Way. However, 
remarkably little is known about the stellar halos of other spiral galaxies, 
and their formation histories.  

Based on the analysis of chemical, kinematic, and structural properties 
of the Galaxy halo stellar populations, two canonical scenarios of halo 
formation and evolution were proposed. The monolithic collapse scenario 
was proposed by Eggen, Lyden-Bell, \& Sandage (1962), in which the stellar 
halo is formed by a rapid collapse of the proto-Galaxy, within a time 
on order of the dynamical timescale ($\sim 10^8\,$yr). 
The second scenario was proposed by Searle \& Zinn (1978) in which the 
halo is formed within a longer timescale ($\sim 10^9\,$yr) by accretion 
of protogalactic fragments that underwent separate pre-enrichment.
The recently identified streams in halos of different galaxies (the Galaxy: 
Yanny et al. 2000; M~31: Ibata et al. 2001; NGC~5128: Peng et al. 2002) give 
credence to the idea that halo populations are affected by infall/accretion 
of dwarfs (Ferguson et al. 2002). 
 
However neither of these scenarios account qualitatively and quantitatively
for the accumulated constraints on the properties of the Milky Way 
stellar halo, and both of them have several difficulties in explaining 
the observed properties of the Galaxy halo stars (Norris \& Ryan 1991; 
Beers \& Sommer-Larsen 1995; Chiba \& Beers 2000); it is yet unclear which
framework may describe accurately the formation and the evolution of the 
Galaxy halo. The failure of these models suggested hybrid scenarios. 
In those scenarios, the flattened distribution of stars at the inner part 
of the Milky Way results from dissipative collapse of large gas clouds 
similar to the Eggen et al. scenario, while the outer Galaxy spheroid 
region originates from dissipationless accretion of dwarf-like fragments 
(Freeman 1996; Carney et al. 1996). 

The main conclusion drawn from the comparison between the Milky Way and 
M~31 (Durrell et al. 2001) halo star properties is that the Galaxy halo 
may not be typical: the halos of spiral galaxies seem to be quite diverse. 
The physical processes thay may regulate this diversity are still very 
unclear (see Bekki \& Chiba 2001 for a discussion of this issue on the 
framework of Cold Dark Matter models). 

A key quantity that helps disentangle the formation history of the 
stellar halo is the metallicity distribution function of galactic halo 
stars. For stellar populations older than about 1 Gyr, the location of 
red giant stars on the color-magnitude diagram is predominantly sensitive 
to metallicity rather than age, so a first-approximation metallicity 
distribution can be constructed through stellar photometry. This approach 
was extensively used, both from ground-based and Hubble Space Telescope 
(HST) deep photometry, to construct the metallicity distribution of 
nearby galaxies. Only a few galaxies beyond the Local Group have measured
metallicity distributions, e.g. NGC~5128 (Harris \& Harris 2002) and
NGC~3379 (Gregg et al. 2004).

In this paper we report the results from the observations of resolved stars 
in halos of a sample of nearby spiral galaxies. The photometry is used to 
construct the metallicity distributions of halo stellar populations. The main 
aim here is to study the global properties of population II halos in normal 
spirals, to seek correlations between galactic properties and halo stellar 
populations and structures.

The layout of this paper is as follows: in \S~\ref{data} we present our 
data set, in \S~\ref{analysis} we present the methodology and the ingredients 
we used to construct the metallicity distribution, and present the properties 
of the observed halo fields, metallicity distributions, and discussions of 
our results. Finally, in \S~\ref{concl} the results of the present work are 
summarized, and their implications for the formation and evolution of galactic
halos are discussed.

\section{ Data}
\label{data}
\subsection{Data reduction and Artificial star experiments}
 
\begin{figure}
\includegraphics[height=3.in]{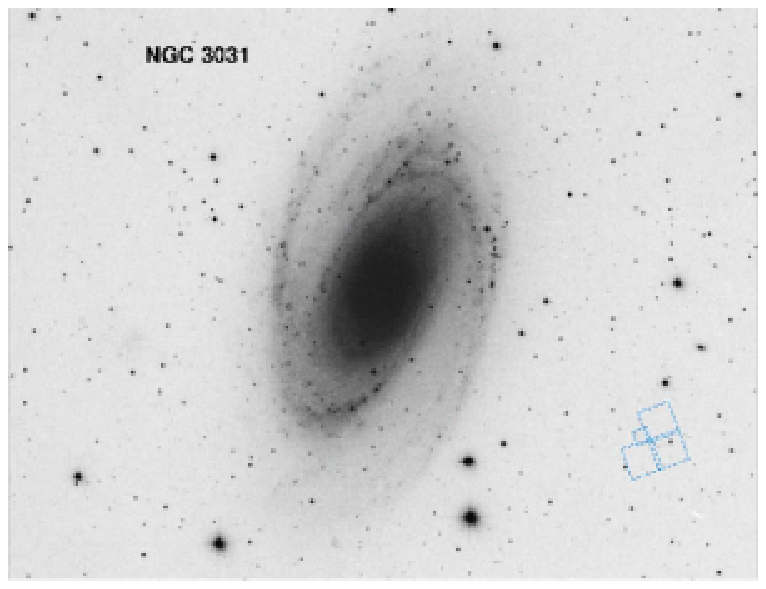}
\includegraphics[height=3.in]{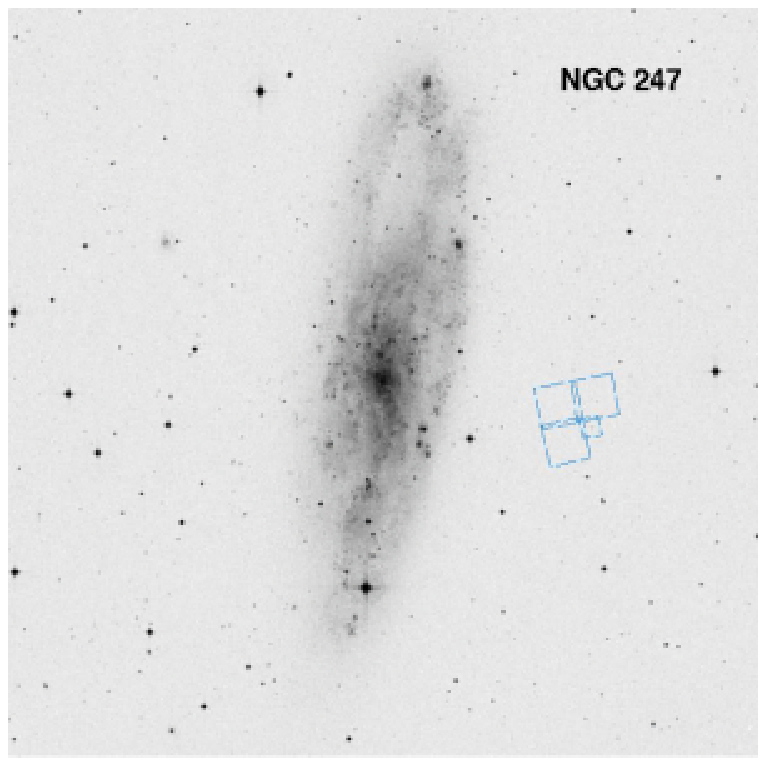}
\includegraphics[height=3.in]{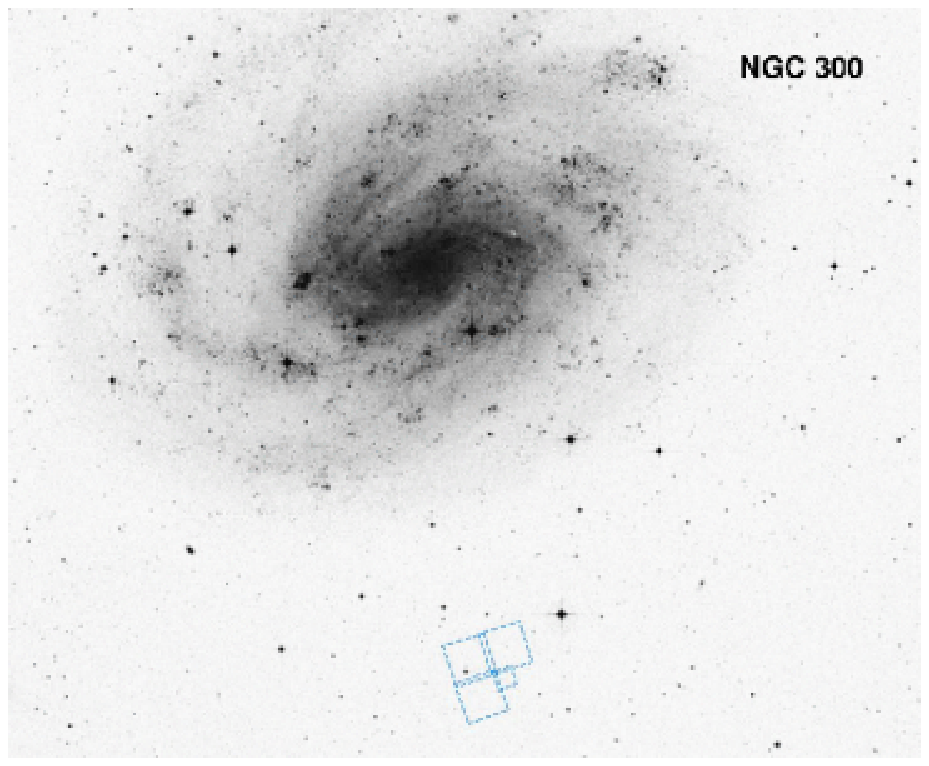}
\includegraphics[height=3.in]{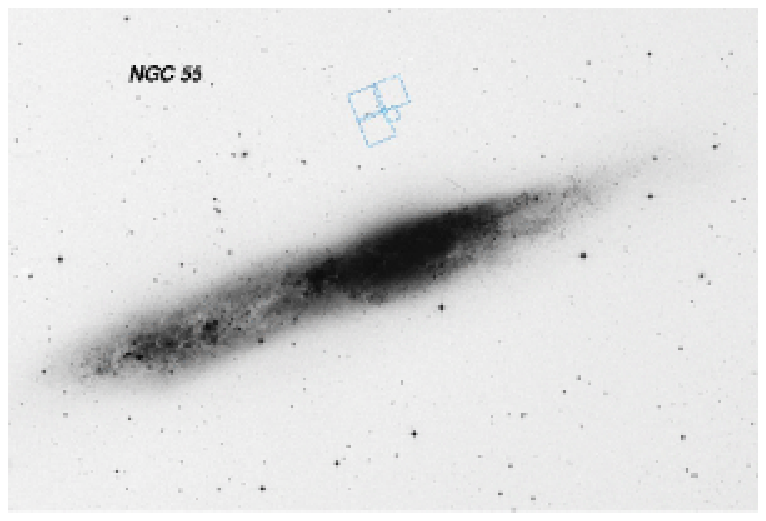}
\caption{HST/WFPC2 footprint of our observations overlayed on 
Digitized Sky Survey images of our sample galaxies that were 
not shown in Mouhcine et al. (2004). }
\label{field_loc}
\end{figure}

The galaxy selection criteria, observations and the reduction techniques 
applied to our sample of highly inclined spiral galaxies have been described 
in Mouhcine et al. (2004a). Here we recall briefly the reduction procedures. 
The galaxies were imaged with the F814W (I) and F606W (V) filters. The location 
of the observed fields that were not discussed in Mouhcine et al. (2004a) 
are shown in Fig.\ref{field_loc}. The observed fields are located so as 
to avoid contamination from the outer bulge and/or the disk, and to sample 
a pure halo stellar population. Exposure times through the F814W filter were 
set to reach a signal-to-noise of 5 in the WF camera for an absolute magnitude 
$M_I = -1$ (for galaxies with $(m-M)_0 < 27$) or $M_I = -2$ for more distant 
ones. The F606W exposure times were set to reach the same S/N at the same 
absolute magnitude for metal-poor RGB stars. The images were reduced using the 
standard HST pipeline, and employing the latest flatfield observations and using 
contemporaneous super-dark reference frames. The dithered frames were combined 
with iterative cosmic ray rejection using software in the stsdas.dither package 
based on the drizzle algorithm of Fruchter and Hook (2002).
The stellar magnitudes were measured through circular apertures with a radius 
of $0.09\arcsec$. These aperture magnitudes were corrected to total magnitudes 
using TinyTim model (Krist 2004) point-spread functions. Aperture-corrections 
and zeropoints are applied separately for each CCD chip. 
The charge-transfer inefficiency has been
corrected using the latest version of the Dolphin (2002) equations.
The instrumental magnitudes were transformed to standard V and I magnitudes
following the prescription of Holtzman et al. (1995). After correcting for
the foreground extinction, we have estimated the magnitudes in the standard
photometric system using the instrumental magnitudes and solving the equations 
relating the WFPC2-to-VI magnitudes using the coefficients tabulated by 
Holtzman et al. (1995).  

As the focus of this paper is to derive the abundance distributions
from a star-by-star analysis, sources of contamination and error
must be understood and controlled.
One of the factors that could influence the color width of the red 
giant branch, and thus affect the measurement of the stellar metallicities, 
are the photometric errors. The photometric errors may introduce a broadening 
of the red giant branch sequence that may be interpreted as indicative of a
physical metallicity distribution function.   Photometric
incompleteness may mask the presence of extremely red stellar populations. 
In order to assess these effects, we carried out a large number of 
artificial star experiments, evaluating the photometric scatter and 
completeness as a function of magnitude. In each 
simulation, stars were added to the original images, with input colors and 
magnitudes following a narrow red giant branch sequence. The locations of 
stars in each frame were chosen randomly, so that over different experiments 
the added stars were uniformly distributed over the whole frame. The images 
were then reduced following the same procedure used for the original frames. 
%
%
%
%
Because the incompleteness varies as a function of I and (V-I) (and on 
the chip), we have mapped out the completeness using linear interpolation 
between the grid points sampled in the artificial star experiments. 
We have used this to assign to each star a completeness fraction. 
In practice, we compute the completeness in terms of the instrumental 
magnitudes and then translate them through the photometric calibration 
equations into the completeness in terms of standand magnitudes for each 
point in the color-magnitude diagram. The depth of the images vary from 
galaxy to galaxy, with $50\%$ completeness limits ranging from absolute
magnitudes $-0.9$ to $-3.5$. To assess the accuracy of our 
photometry, we calculate the difference between the input and recovered 
magnitude for each magnitude bin. The photometric errors at an absolute 
magnitude ${\rm M_I = -3}$ range from $\sim\,0.03$ mag in the best case 
(NGC~55) to $\sim\,0.15$ mag in the worst case (NGC~4258).

For the purpose of deriving the metallicity distribution for stellar halos, 
the completeness effects may be large for extremely red and bright giant 
stars at the upper right corner of a color-magnitude diagram near the 
completeness cutoff determined by the V-filter; those stars are likely 
to have solar or super-solar abundances. For nearly all sample galaxies, 
apart from NGC~4258, the $50\%$ completeness limit falls well below the 
brightest end of the red giant branch. In our analysis of the metallicity 
distribution, we include the effects of incompleteness as far as the data 
allow. 
Contamination by foreground Galactic stars may be important, especially 
for the NGC~4945 halo field because of its low Galactic latitude and the 
deep photometry.  To estimate Galactic foreground contamination, we used 
the Besan\c{c}on group model of stellar population synthesis of the Galaxy
available through the Web$\footnote{http://bison.obs-besancon.fr/modele/}$ 
(Robin et al. 2003), simulating the total number and optical photometry of
foreground stars. We found that Galactic field star contamination contributes 
only $\sim\,3\%$ at the maximum, in the case of NGC~4945, within the magnitude 
ranges used to construct the metallicity distribution functions and 
${\rm -2\le\,(V-I)\le\,5}$. 
A few faint background galaxies may have misclassified as stars. An upper 
limit to the contamination can be estimated from the Hubble Deep Fields
(catalogs described in  Casertano et al., 2000), which have the same area 
as the fields we have observed.
In the HDF-N we find 10 sources with $23 < F814W < 26$ and $FWHM < 0.2$
arcsec. In the HDF-S there are 22 such sources. Some of these are galactic 
foreground stars. Thus the level of compact-source contamination is entirely 
negligible.

\subsection{Distances and Reddening}

The morphological type, disk inclination, integrated  magnitude,
foreground extinction, and true distance modulus of each galaxy in the 
sample are listed in Table\,\ref{gal_prop}. 
Extinction- and inclination-corrected V-band magnitudes come from RC3. 
No significant internal extinction is expected to affect the halo stellar 
populations, as halo regions are most likely dust free. The reddening 
toward the halo fields was estimated using the all-sky map of Schlegel 
et al. (1998). We adopt the relations ${\rm A_{F606W}=2.677\times E(B-V)}$ 
and ${\rm A_{F814W}=1.815\times E(B-V)}$ for the used HST filters. We have 
neglected the effect of any possible differential reddening across our 
fields and along the line of sight. Indeed, the differential reddening due 
to the Galactic foreground across the WFPC2 fields is likely to be small. 
The locations of the observed halo field are shown Fig.\ref{field_loc}, 
on the Digitized Sky Survey image of the galaxies that were not discussed 
in paper I, i.e., NGC~55, NGC~247, NC~300, and NGC~3031.

As demonstrated extensively in the literature, the tip of the red giant 
branch is a useful distance indicator (Lee et al. 1993). The I-band 
luminosity function can be used to identify the magnitude level of the 
tip of the red giant branch, and thus the distance modulus to our sample 
galaxies. Mouhcine et al. (2004a) have used both the edge-detection 
algorithm and the maximum likelihood analysis to estimate the tip of the 
red giant branch magnitude for four galaxies in our sample, i.e., NGC~253,
NGC~4244, NGC~4258, and NGC~4945 (see Mouhcine et al. 2004a for detailed
discussion). For the other four galaxies in our sample, i.e. NGC~55, 
NGC~247, NGC~300, and NGC~3031, the numbers of stars at the bright 
end of the red giant branch sequences are too small. Consequently, it 
is difficult to locate reliably the discontinuities in red giant star 
luminosity functions, and then to measure accurately the distance modulus 
to those galaxies (Madore \& Freedman 1995). For those galaxies, we rely 
on published distances. NGC3031 (M~81, is a Sb galaxy with similar 
properties to M~31, with an inclination of $60^{\circ}$; it is the nearest 
bright galaxy with both a dynamically known black hole and nuclear activity 
(Ho, Filippenko, \& Sargent 1996). 
The association of galaxies around M~81, i.e., M\,82 (NGC~3034), NGC~3077, 
and various dwarf spheroidal and irregular galaxies, is one of the nearest 
prominent groups in the vicinity of the Local Group. NGC~3031 has been 
observed as a part of the HST Key project (Freedman et al. 1994). 
Ferrarese et al. (2000) published a Cepheid-based distance modulus to 
NGC~3031 of $\mu=27.8 \pm 0.06$. 
The other three galaxies are all members of the Sculptor group. 
Graham (1982) suggest that this small group is significantly extended 
along the line of sight, and a range of $\sim 0.7$ in $(m-M)_{\circ}$ 
may be expected. NGC~300 is an SA(s)d galaxy, with an inclination of 
$46^{\circ}$. Freedman et al. (2001) have published a Cepheid-based 
distance of $2.0\pm 0.1$ Mpc ($(m-M)=26.53$). 
NGC~55 is an Sc galaxy that is almost edge-on, with an inclination of 
$90^{\circ}$. The distance to NGC~55 is poorly known. 
Puche \& Carignan (1988) have compiled different distance measurements 
to the galaxy, with a mean distance modulus of $(m-M)=26.11$. 
Using the I-band luminosity function of the carbon star population in 
the galaxy from Pritchet et al. (1987), the calibrated mean absolute 
I-band magnitude of carbon stars from Richer (1981), and using the long 
distance scale to the Large Magellanic Clouds, the distance modulus to 
NGC~55 of $(m-M)_o=25.66$ (see Richer et al. 1985 for a discussion of
empirical evidence for using carbon stars as a distance indicator and 
Mouhcine \& Lan\c{c}on 2003 for a theoretical discussion of the issue). 
This is consistent with Graham's conclusion that NGC~55 lies in front 
of NGC~253 (Graham 1982). However, the Pritchet et al. (1987) survey 
of carbon stars in the galaxy suffers from a severe incompleteness. 
In our analysis, we will use the distance modulus from Puche \& Carignan 
(1988). 
It is worth mentioning that the uncertainties in the distance 
modulus to NGC~55 have little effect on the analysis presented later 
in the paper.
NGC~247 is an Sc(s)III-IV galaxy, almost edge-on galaxy (with an inclination 
of $80^{\circ}$). NGC 247 has the most uncertain distance modulus estimate. 
Davidge \& Courteau (2002) have used the Tully-Fisher relation, as calibrated 
by Sakai et al. (2000), with published integrated photometry and line widths 
from Pierce \& Tully (1992), to derive a galaxy distance modulus of 
$(m-M)_{\circ}=27.3$. The derived distance is consistent with NGC~247 being 
a member of the Sculptor group at its far side, and with its location on the 
sky close to NGC~253.

\section{ Analysis}
\label{analysis}

\subsection{Color-magnitude diagrams}
\label{cmd_sample}

\begin{figure}
\includegraphics[height=3.5in]{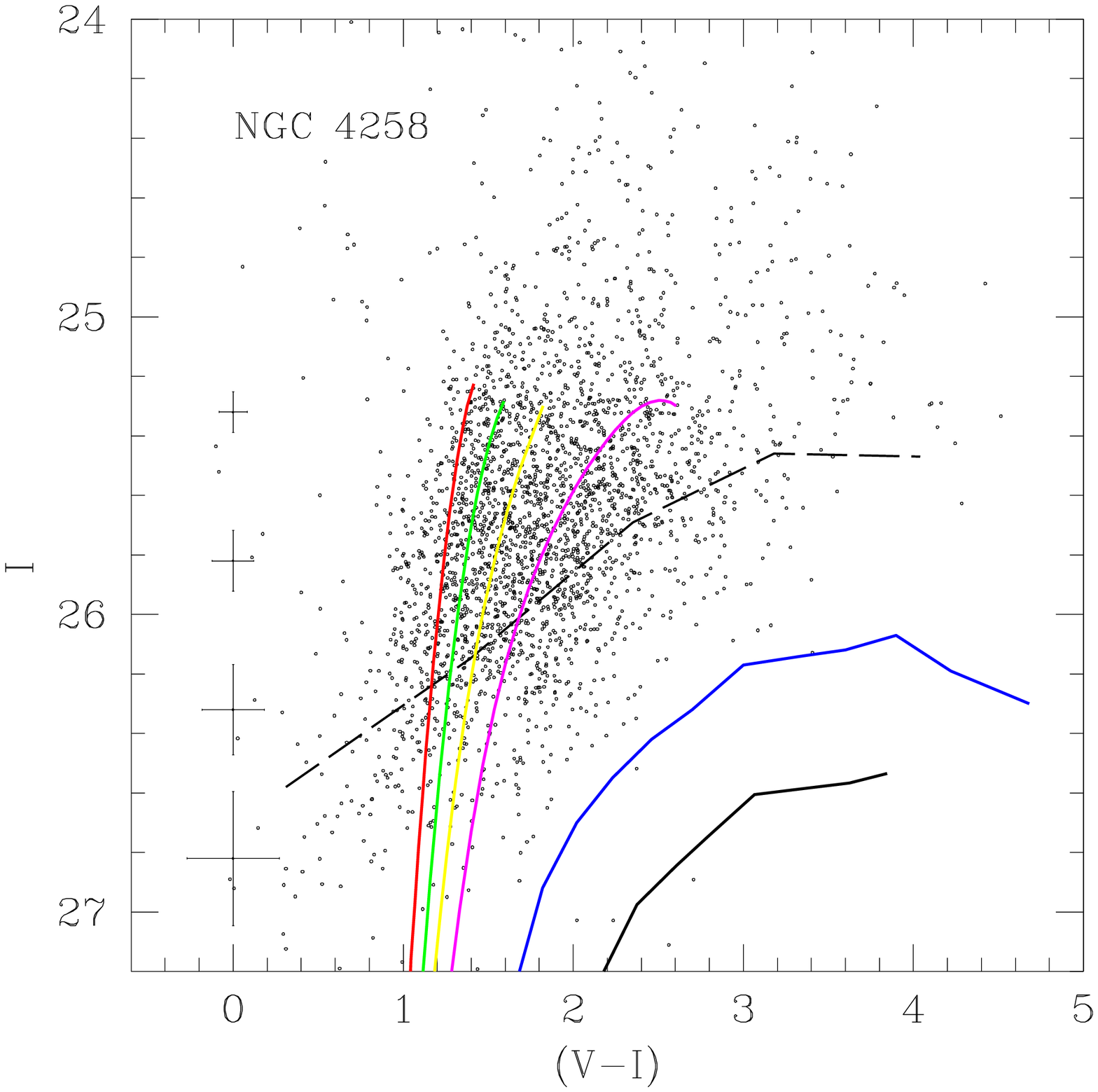}
\includegraphics[height=3.5in]{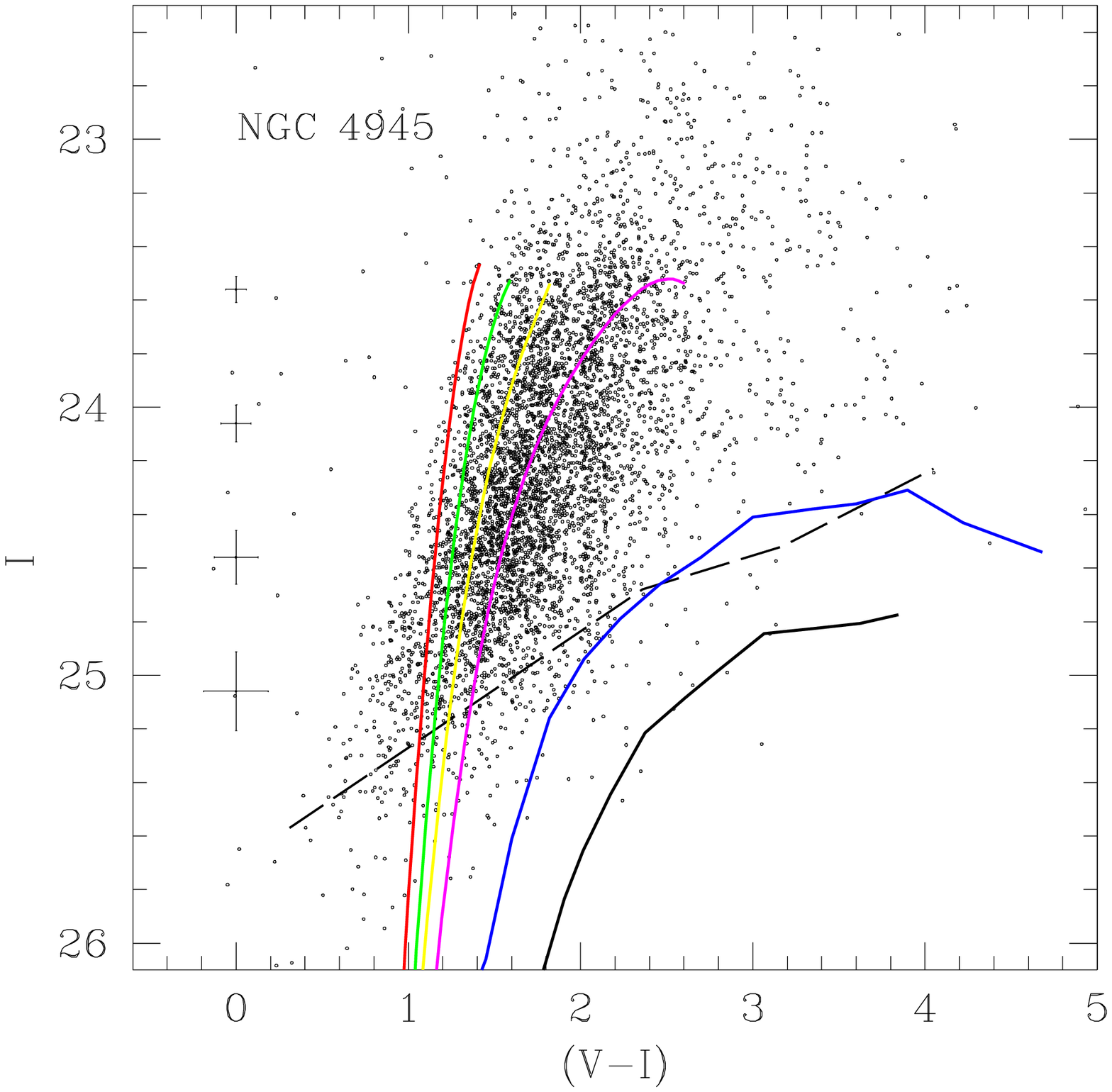}
\includegraphics[height=3.5in]{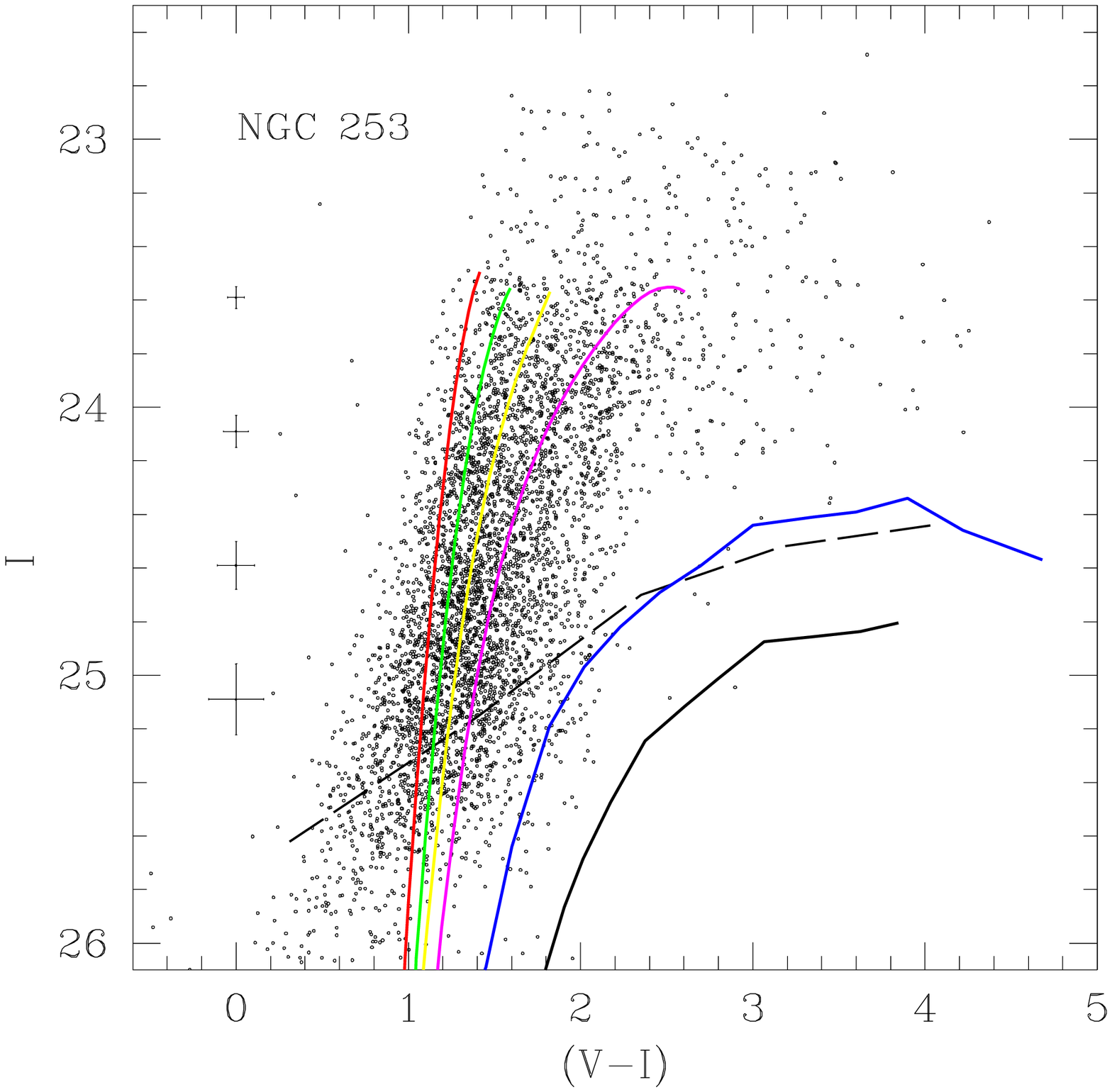}
\includegraphics[height=3.5in]{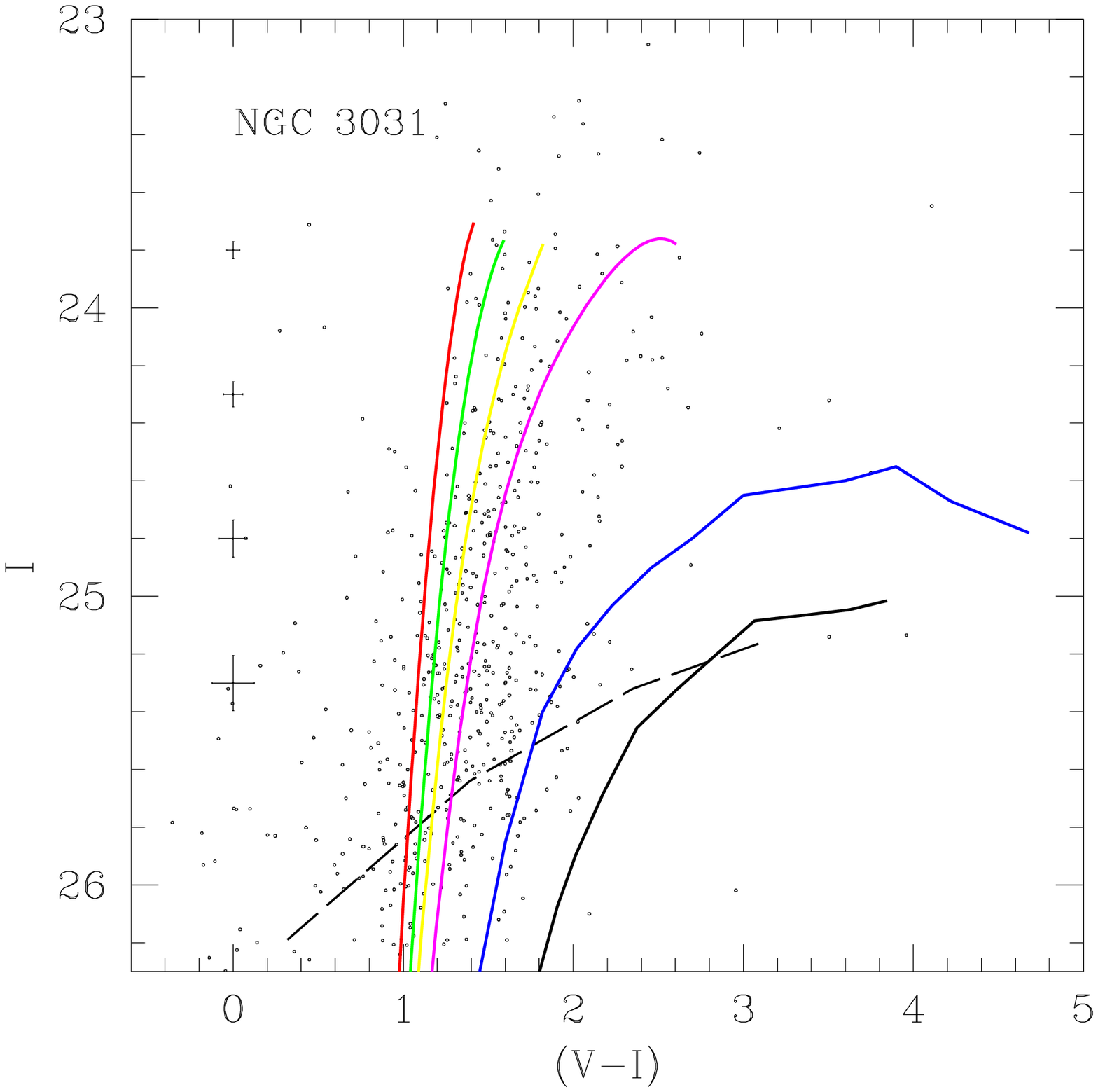}
\caption{Color-Magnitude diagrams of the observed halo fields, with 
fudicial lines from Galactic globular clusters superimposed. These are 
M~15 ([Fe/H]=-2.2), M~2 ([Fe/H]=-1.6), NGC~1851 ([Fe/H]=-1.3), 47~Tuc 
([Fe/H]=-0.7), NGC~6553 ([Fe/H]=-0.25), and NGC~6791 ([Fe/H]=0.2).
The plotted error bars denote the errors for objects with (V-I)=1.
The dashed lines show the 50\% detection completeness levels.
The width of the red giant branch shows a large variety, indicating 
differences in the abundance distribution of spiral halo stellar 
populations; halos of spiral galaxies may be composed entirely by 
metal-poor stars, or be comoposed of a mixture of low to intermediate 
metallicity stellar populations.}
\label{galcmd1}
\end{figure}

\begin{figure}
\includegraphics[height=3.5in]{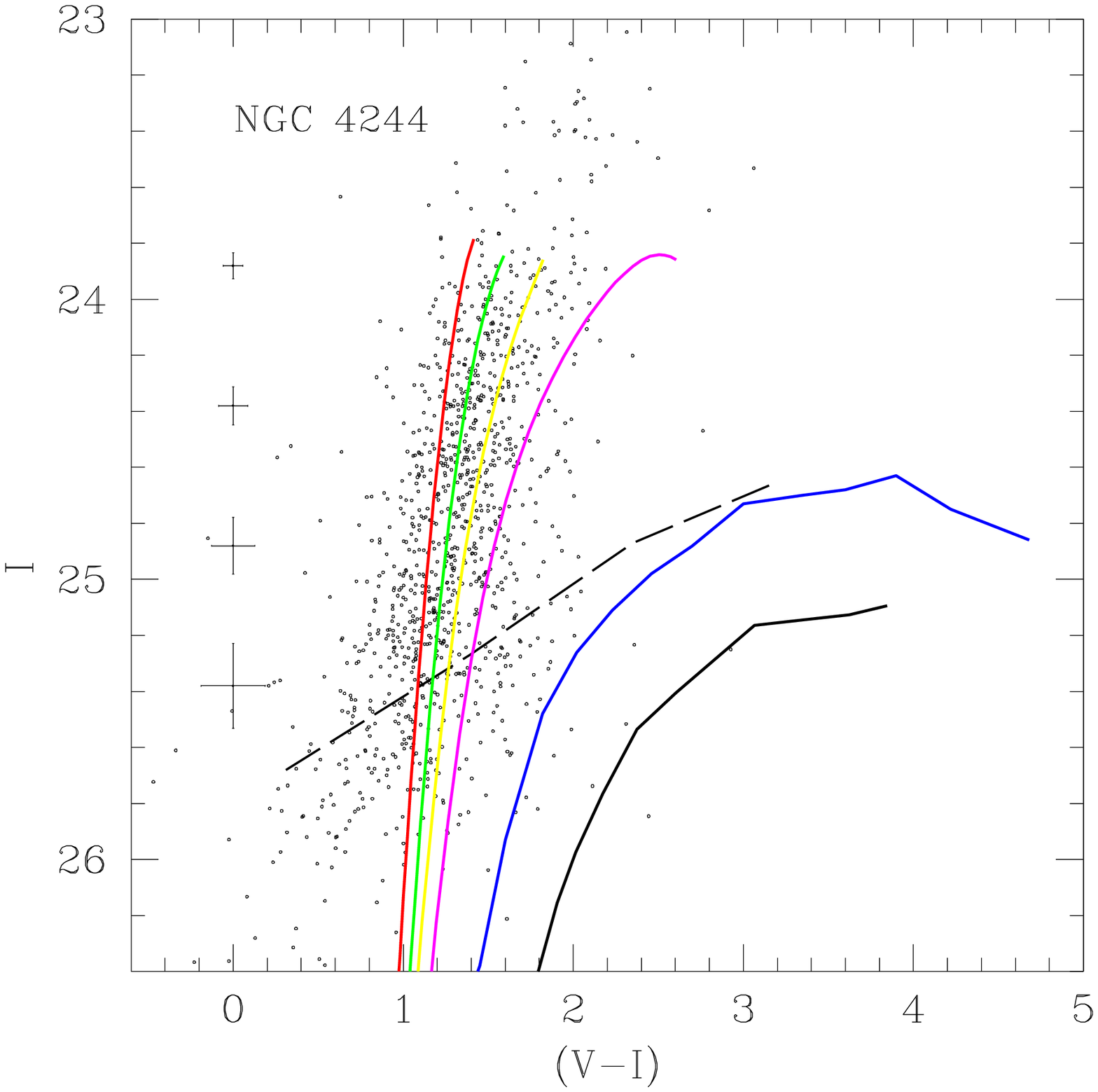}
\includegraphics[height=3.5in]{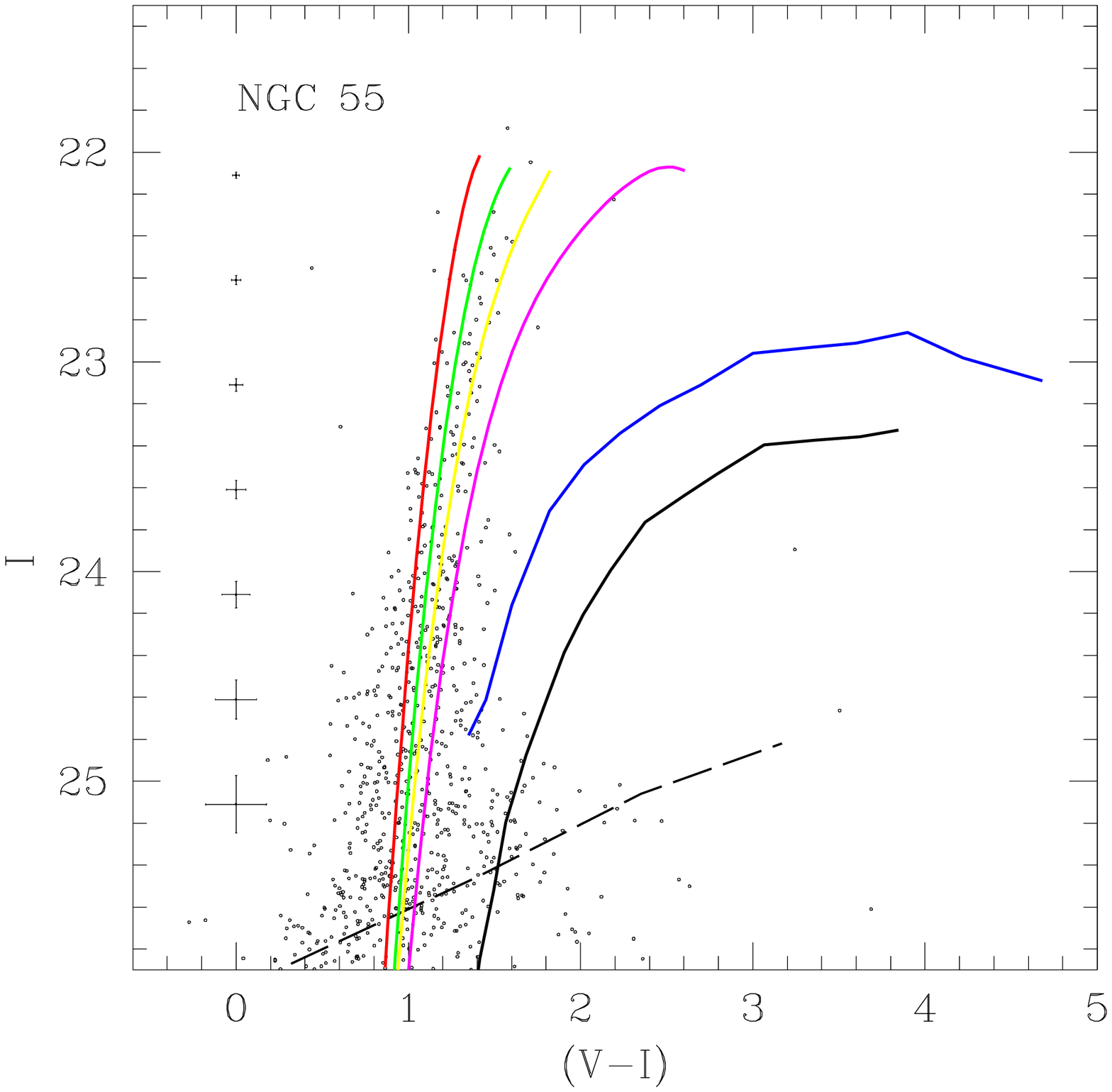}
\includegraphics[height=3.5in]{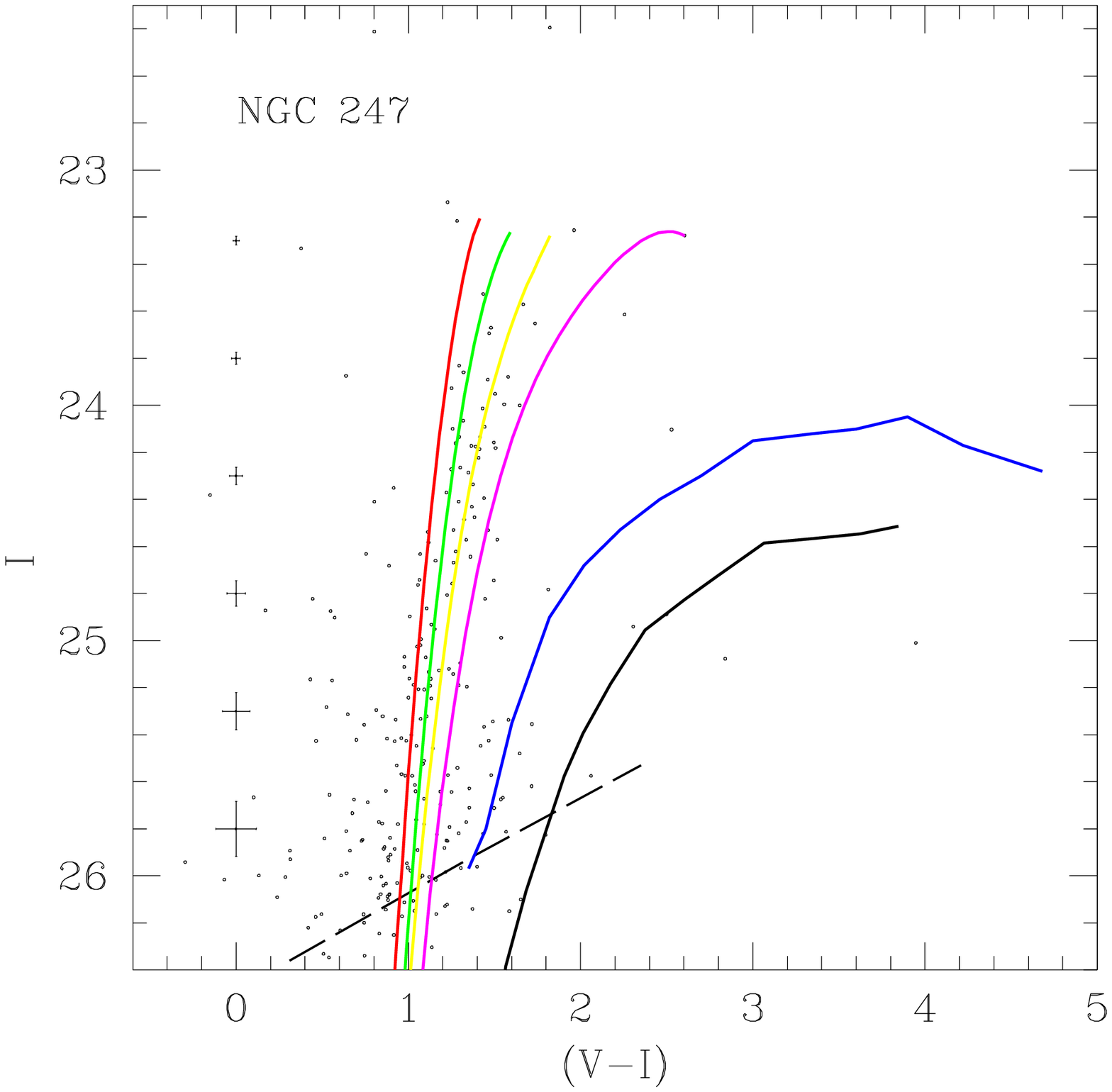}
\includegraphics[height=3.5in]{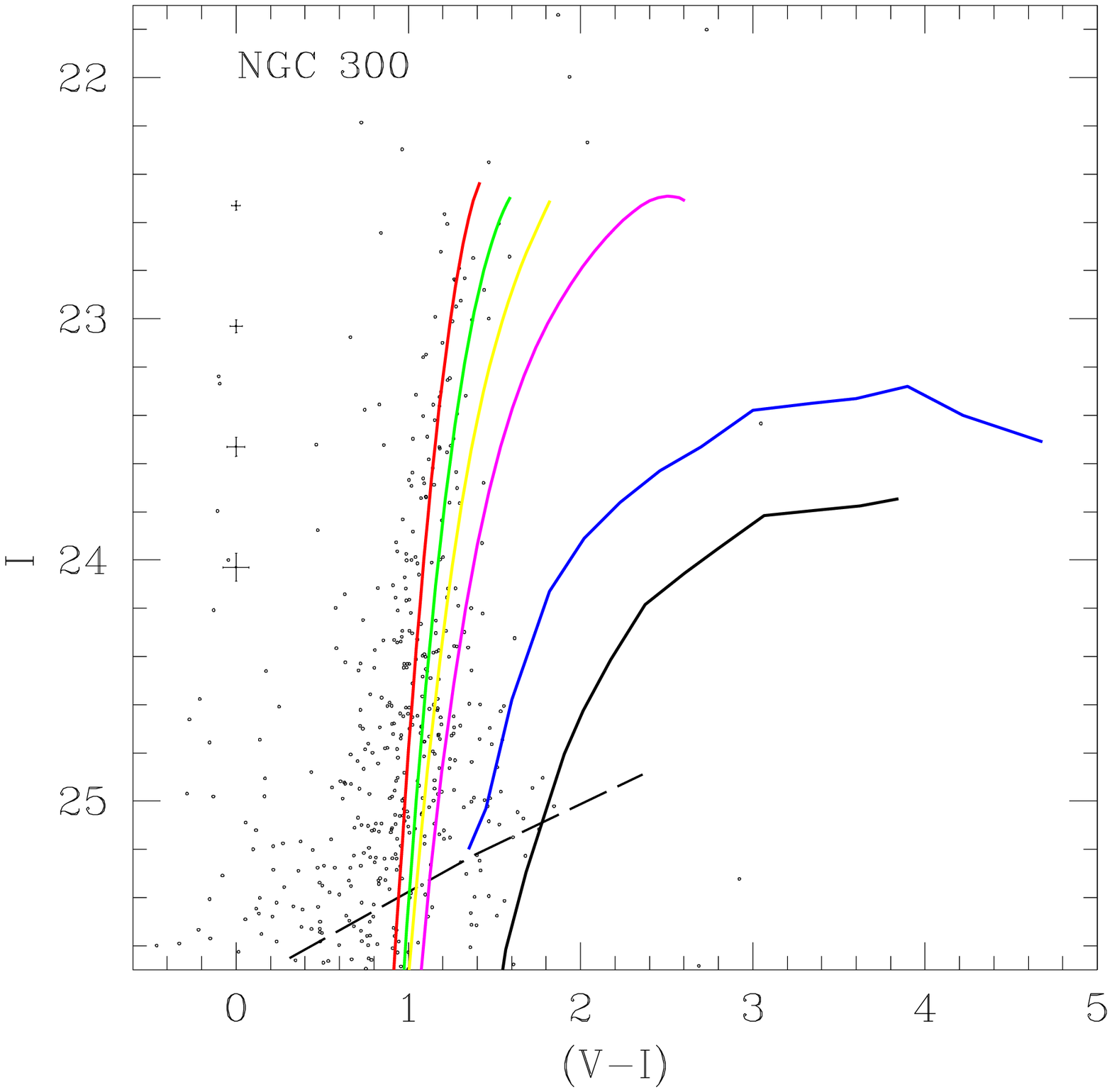}
\caption{Same as in Fig.\,\ref{galcmd1} except for NGC~4244, NGC~55, 
NGC~247, and NGC~300}
\label{galcmd2}
\end{figure}

Fig.\,\ref{galcmd1} and Fig.\,\ref{galcmd2} shows the reddening-corrected 
and distance-corrected color-magnitude diagrams of the sample galaxies. 
The most prominent features are the well-populated red giant branches.
The color-magnitude diagram contains no indication of significant young 
or intermediate-age stellar populations in the halos of these spiral galaxies. 
There are few stars in the region where one may expect to see early-type 
stars; in addition, above the first-ascent giant stars there is a negligible 
number of stars that may identified as bright asymptotic giant branch stars, 
associated with intermediate age populations. A firm conclusion about the 
presence or lack of an intermediate age population would require much deeper 
photometry, as shown recently by Brown et al. (2003) for the halo of M31.

As a first indication of the range of metallicities across the red giant 
branch, we have overplotted the loci of the observed red giant branch 
sequences for standard Milky Way globular clusters of different metallicities 
which encompass the majority of RGB stars for each galaxy. These lines are 
for the observed red giant branch sequences of M~15 ([Fe/H]=-2.2), M~2 
([Fe/H]=-1.6), NGC~1851 ([Fe/H]=-1.3), and 47~Tuc ([Fe/H]=-0.71) from 
Da Costa \& Armandroff (1990), the metal-rich bulge globular cluster 
NGC~6553 ([Fe/H]=-0.25) from Segar et al. (1999), and the old disk open 
cluster NGC~6791 ([Fe/H]=0.2) from Garnavich et al. (1994).   
While we cannot rule out an age spread in these galaxies, the color spread 
observed in the data is due primarily to a spread in metallicity. 
Theoretical isochrones show that metallicity has a much larger effect than 
age on RGB color.  For example, at ${\rm M_I = -3}$ on the red giant branch, 
a 0.2~mag shift in ${\rm (V-I)}$ to the blue can be achieved by a relatively
large decrease in age from 13 to 6 Gyr or a relatively small decrease in 
[Fe/H] from $-1$ to $-1.2$. In the analysis to follow we shall assume that 
halo stellar populations have a fixed globular cluster-like age. Mean 
metallicities will be slightly higher if the ages are lower.
>From the comparison in Fig.\,\ref{galcmd1} and Fig.\,\ref{galcmd2}, it is 
clear that all spiral halos contain a metal-poor stellar component extending 
down to $[Fe/H]\sim -2$, however only bright galaxies, i.e., 
${\rm M_{V,\circ}\sim -20}$, tend to have a stellar component that extends 
to high metallicities, even higher than the metallicity of the most 
metal-rich halo globular cluster 47~Tuc.
With the HST, the efforts of investigating stellar halo properties have 
extended beyond the Local Group to include the Cen A group (e.g., Harris 
et al. 1999) and the giant elliptical NGC~3379 in the Leo I Group (Gregg 
et al. 2004). The color-magnitude diagrams of the halo of spiral galaxies 
show that the overall color-magnitude diagram morphologies of the halo 
stellar populations are similar, showing the same old red giant branch 
population that ends abruptly at the tip of the sequence.

The morphology of the color-magnitude diagrams demonstrates that the stellar 
populations in the halos of spiral galaxies are diverse, and can range from 
a very low metallicity of $\sim -2$ dex up to supra-Solar. The color-magnitude 
diagrams of NGC~4258, NGC~253, and NGC~4945 show a large color span and are 
similar to the color-magnitude diagrams of the halo of M~31 (Durrell et al.
2001, 2004) and NGC~5128 (Harris \& Harris 2000, 2002). 
NGC~3031 and NGC~4244 show a mixture of low-to-intermediate metallicities, 
and have similar morphologies to the color-magnitude diagram of the outer 
regions M~33 (e.g., Tiede et al. 2004). 
The location of the color-magnitude diagram blue edge is similar to what 
is observed for the three bright galaxies, suggesting that their original 
halo gas started at similar chemical compositions, but they have undergone 
slower chemical enrichment which has reached a smaller metallicity spread. 
The faintest galaxies in our sample---NGC~55, NGC~247, and NGC~300--- are 
composed mainly of metal-poor stars, and are similar to the dwarf spheroidal 
M~81-BK5N in the M~81 group (Caldwell et al. 1998).

\subsection{ Metallicity distribution functions}
\label{zdist}

\subsubsection{ Calibration of metallicity measurements}
\label{calib}

To derive the metallicity distribution function for the galaxy sample, we 
convert the observed stellar photometry to metallicity on a star-by-star 
basis. To do so, we superimpose a fiducial grid of RGB tracks on the 
color-magnitude diagrams, and interpolate between them to derive an 
estimate of a star's metallicity (e.g., Holland et al. 1996; Harris et al.
1999). Such a procedure stands implicitly on the assumption that all of 
the halo stellar population has a uniformly globular cluster-like age. 
The shape of the metallicity distribution will change only slightly if 
the age distribution is narrow. The random errors in the color measurements 
are significantly smaller than the observed color spread at the bright 
red giant branch end, used to derive the metallicity distribution function. 

A bias may be caused by the presence of early-asymptotic giant branch stars
(E-AGB), which run almost parallel to the red giant branch with bluer colors, 
and hence may be misclassified as stars on the red giant evolutionary phase. 
The inclusion of these stars may bias the metallicity distribution function 
toward an excess of metal-poor stars. However, Padova group stellar 
evolutionary models (e.g., Fagotto et al. 1994) indicate that the lifetime 
ratio of E-AGB phase to RGB phase, directly related to the number ratio of 
E-AGB stars to RGB stars, is 15--20\%, suggesting that the bias exerted by
E-AGB stars should not be severe (see also Harris et al. 1999 and Durrell
et al. 2001). Note that a fraction of metal-poor E-AGB 
stars are rejected from the sample of stars used to build the metallicity 
distributions as they are bluer than the colors of the most metal-poor 
track in the grid. Another bias source is the possibility of ages 
considerably younger than globular clusters. Brown et al. (2003) 
have shown that the halo of M~31 contains a significant intermediate-age, 
metal-rich stellar population, as well as the classical globular cluster 
age-like metal-poor stellar population. 
Lacking information about the age distribution of stars 
in our sample, we shall simply assume that they are all old and of uniform 
age. While we do not relax this assumption in this paper, it is worth noting 
here that a fairly plausible age distribution, wherein metal-rich stars are 
significantly younger than metal poor stars, will tend to reduce the color 
spread in the RGB. We shall return to this point in \S~\ref{mdf_prop}.

To generate the metallicity distribution of halo stars we prefer using 
theoretical red giant branch sequences over empirical cluster fiducials. 
Even with the remaining uncertainties, such as the effect of convection 
on the evolution of RGB stars, or the temperature scale of giant stars, 
the state of the theoretical modeling of giant stars should be accurate 
enough to achieve the goal we seek. Harris \& Harris (2000) have shown 
that the coarness of the observed cluster fiducial grid may cause 
artifacts in the metallicity distribution function that disappear once 
a densely spaced grid in metallicity is used.
We use VandenBerg et al. models (2000) of red giant tracks for 
${\rm 0.8\,M_{\odot}}$ stars and covering the abundance range from 
${\rm [Fe/H]=-2.314}$ to ${\rm [Fe/H]=-0.397}$, approximately in steps of 
0.1 dex. All the models we used assume that the stars are $\alpha-$enhanced, 
i.e., [$\alpha$/Fe]=0.3, to take account of recent results on the abundance 
analysis of Galactic halo stars (McWilliam 1997, Gratton et al. 2000), and 
in the globular cluster of nearby early-type galaxies (Larsen et al. 2002). 
The metallicity is defined as ${\rm [M/H]=[Fe/H]+[\alpha/Fe]=\log(Z/Z_{\odot})}$.
The grid has been compared and well calibrated against the observed loci of 
standard globular cluster fiducial sequences (Bergbusch \& VandenBerg 2001;
Harris \& Harris 2000). 
We attempted to use the semi-empirical red giant grid constructed by 
Saviane et al. (2000), but realized that the metallicity range covered 
by this grid is not sufficient to cover the range of stars observed in 
halos of some galaxies in our sample. Note there is good agreement 
between the Saviane et al. (2000) calibration and the grid we are using, 
over the metallicity range common to both red giant sets. 

For bright galaxies in the sample, a number of halo stars are located 
beyond the most metal-rich track of the theoretical grid, suggesting 
the presence of metal-rich stellar populations in the halo of these 
galaxies. To cover these stars, and as the theoretical stellar 
evolutionary tracks of super-solar metallicity stars are still of poor 
quality, we add to this grid the observed red giant branch fiducial of 
the old metal-rich disk open cluster NGC~6791. The metallicity of the 
cluster is thought to lie in the range ${\rm [Fe/H]=0.16-0.44 dex}$ 
(Taylor 2001, and reference therein), with no $\alpha-$element 
enhancement (Peterson \& Green 1998, Chaboyer et al. 1999). 
The estimate of the metal-rich end of the metallicity distribution function 
suffers from both the poor knowledge of the metal-rich star evolutionary 
patterns, and from the photometric incompleteness affecting preferentially 
the regions in the color-magnitude diagrams where these stars are located. 
The calibration of the grid at the metal-rich end is the most uncertain part 
of the analysis of our data set.

To evaluate the stellar metallicities, we proceed as follows. For each 
observed stellar distance- and reddening-corrected I-band magnitude, we 
$(i)$ construct a relationship between the color (V-I)$_o$ of fiducial 
tracks at this magnitude, and their metallicities [M/H], $(ii)$ then we 
fit a non-linear function to the relationship, $(iii)$ we use the fit 
to the relationship between (V-I)$_o$ and [M/H] to convert the observed 
reddening-corrected stellar color to a metal abundance. This was repeated
for all stars in the color-magnitude diagrams lying in the magnitude ranges 
used to derive the metallicity distributions. Similar procedures were used 
in the literature (e.g., Holland et al. 1996; Grillmair et al. 1996; 
Harris \& Harris 2000).

In the brightest $\sim 0.5$ magnitude of the RGB, the V band blanketing 
causes the V, I distribution of the metal-rich stars turn over, making 
the interpolation procedure rather uncertain. On the other hand, the 
faint end of the red giant branch is affected by large photometric 
errors, and will scatter more stars to bluer color, which may bias the 
resultant metallicity distribution to be more metal poor. 
To construct the metallicity distribution function, we therefore decided 
to avoid both the bright and the faint end of the observed red giant 
sequence, and restrict our analysis to stars in the magnitude range that 
minimize the effects of both the photometric errors, incompleteness, and 
the evolutionary track curvature. The magnitude ranges used to estimate 
the metallicity distribution function for each galaxy are indicated in 
the panels of Fig.~\ref{mdf_mdl1} and Fig.~\ref{mdf_mdl2}.
To overcome the problem of track curvature at the high metallicity end, 
Harris \& Harris (2000, 2002) and Durrell et al. (2001) have followed 
similar procedures, but interpolated in the ${\rm M_{bol} vs. 
(V-I)_{\circ}}$ plane. In principle, using stellar magnitudes, colors, 
and bolometric corrections, the bolometric magnitude can be estimated. 
However, the bolometric corrections depend on the stellar metallicity, 
and this is exactly what one is looking for. This circularity affects 
the self-consistency of such an approach.  
Our procedure is valid as long as stars are within the evolutionary RGB 
track grid, but we decided to include stars that are one $\sigma_{V-I}$ 
bluer, at a given reddening-corrected I-band magnitude, than the lowest 
metallicity track, as well as for stars redder than the highest metallicity 
fiducial track. 

\subsubsection{ Results for the metallicity distribution function}
\label{mdf_prop}

\begin{figure}
\includegraphics[height=3.5in]{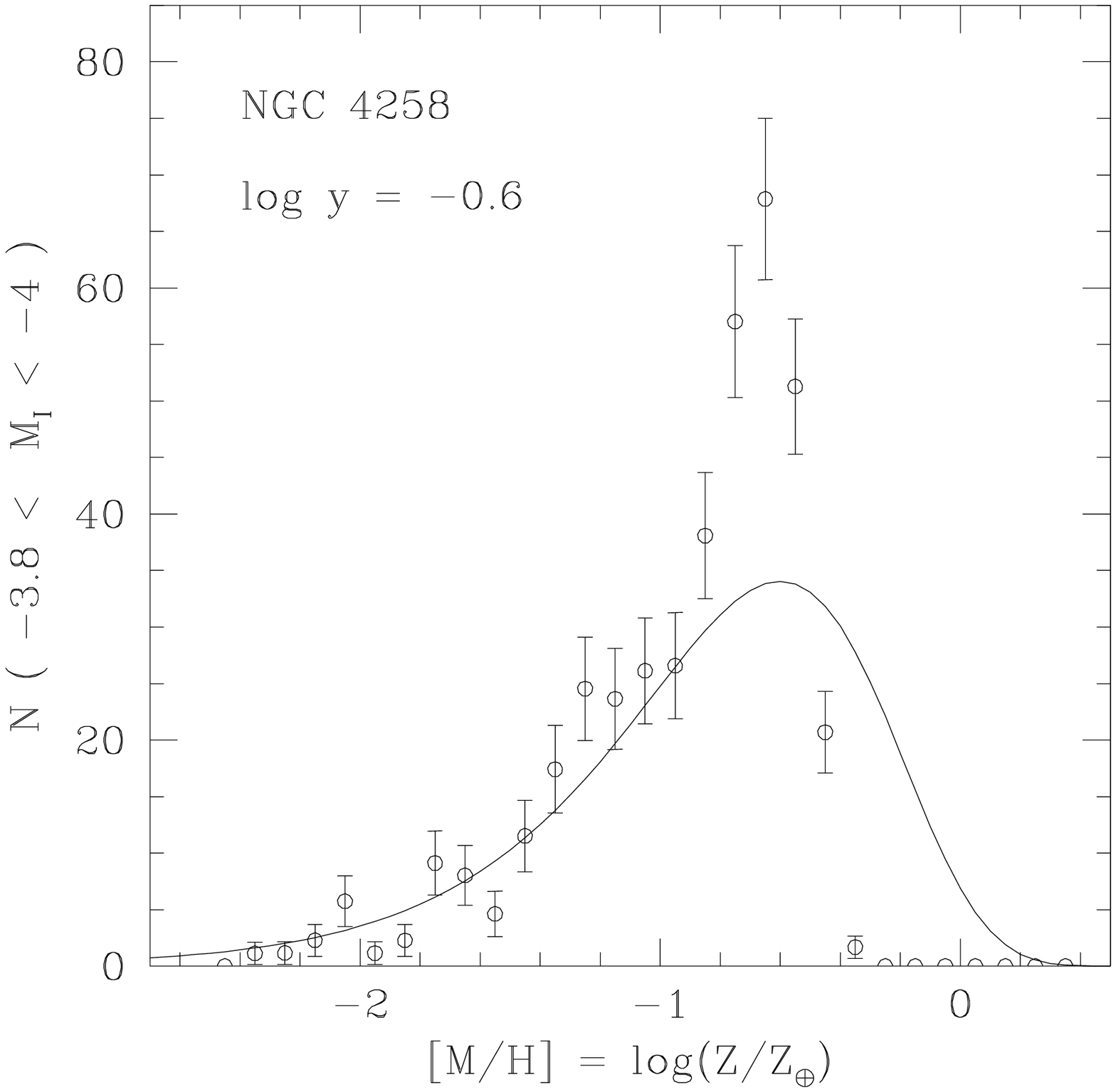}
\includegraphics[height=3.5in]{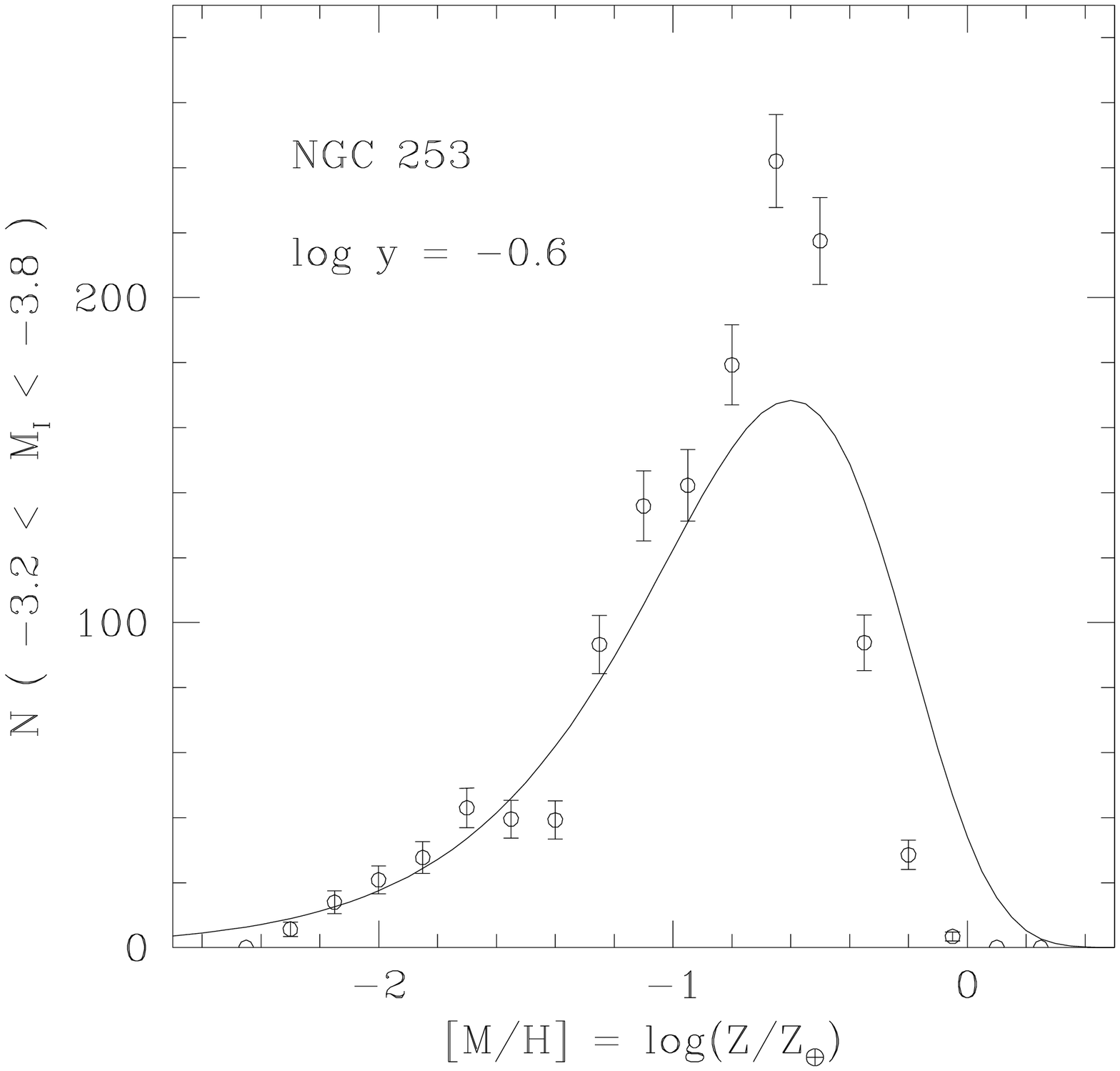}
\includegraphics[height=3.5in]{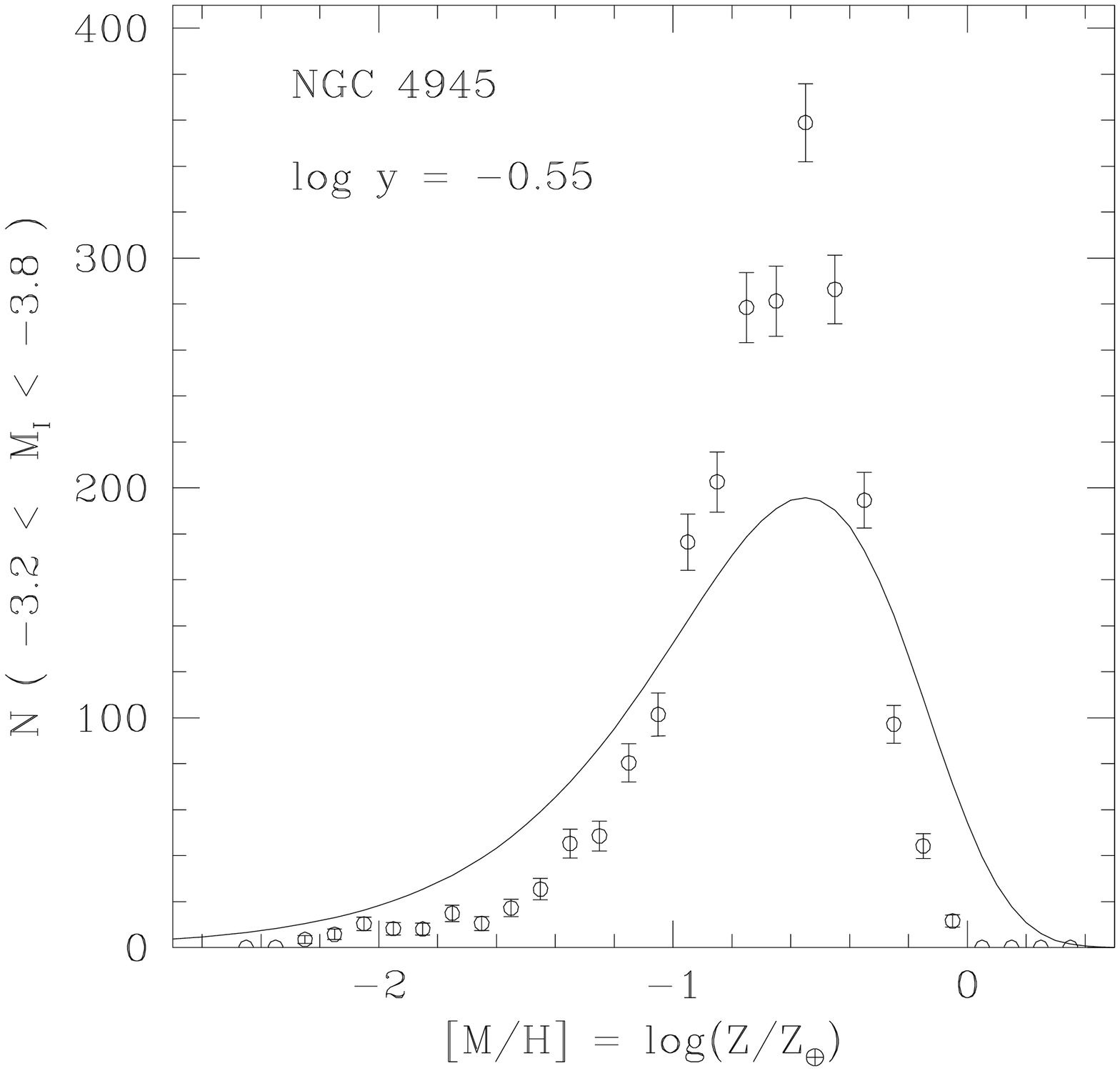}
\includegraphics[height=3.5in]{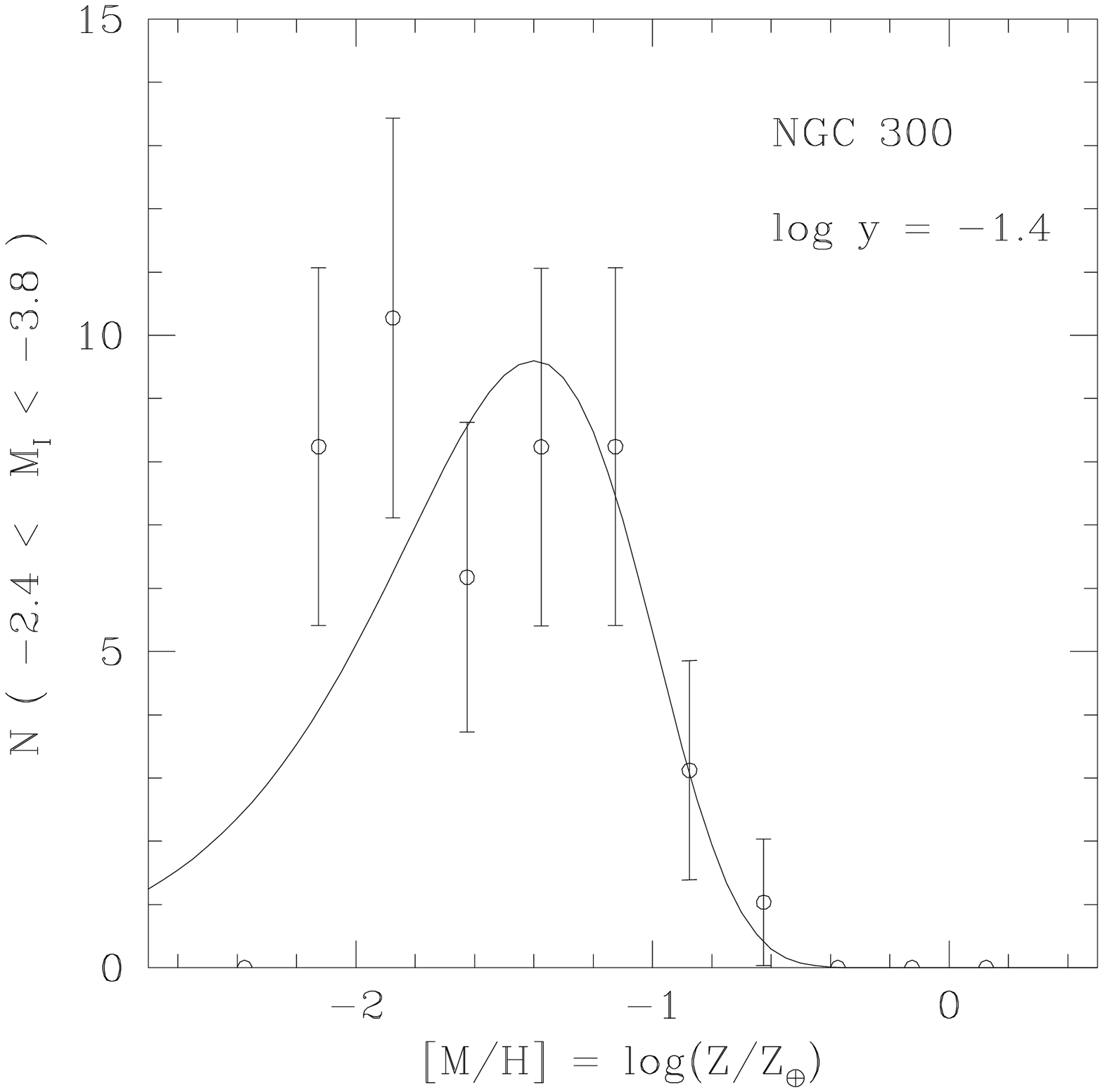}
\caption{Comparison between the metallicity the metallicity
distribution of halo giant stars and the prediction of 
the simple chemical evolution model (closed-box model)
with the indicated value of the yield ($y=<Z>$) for 
NGC~4258, NGC~253, NGC~4945, and NGC~3031. The model 
distributions have been scaled by the total area under 
the data histogram.}
\label{mdf_mdl1}
\end{figure}

\begin{figure}
\includegraphics[height=3.5in]{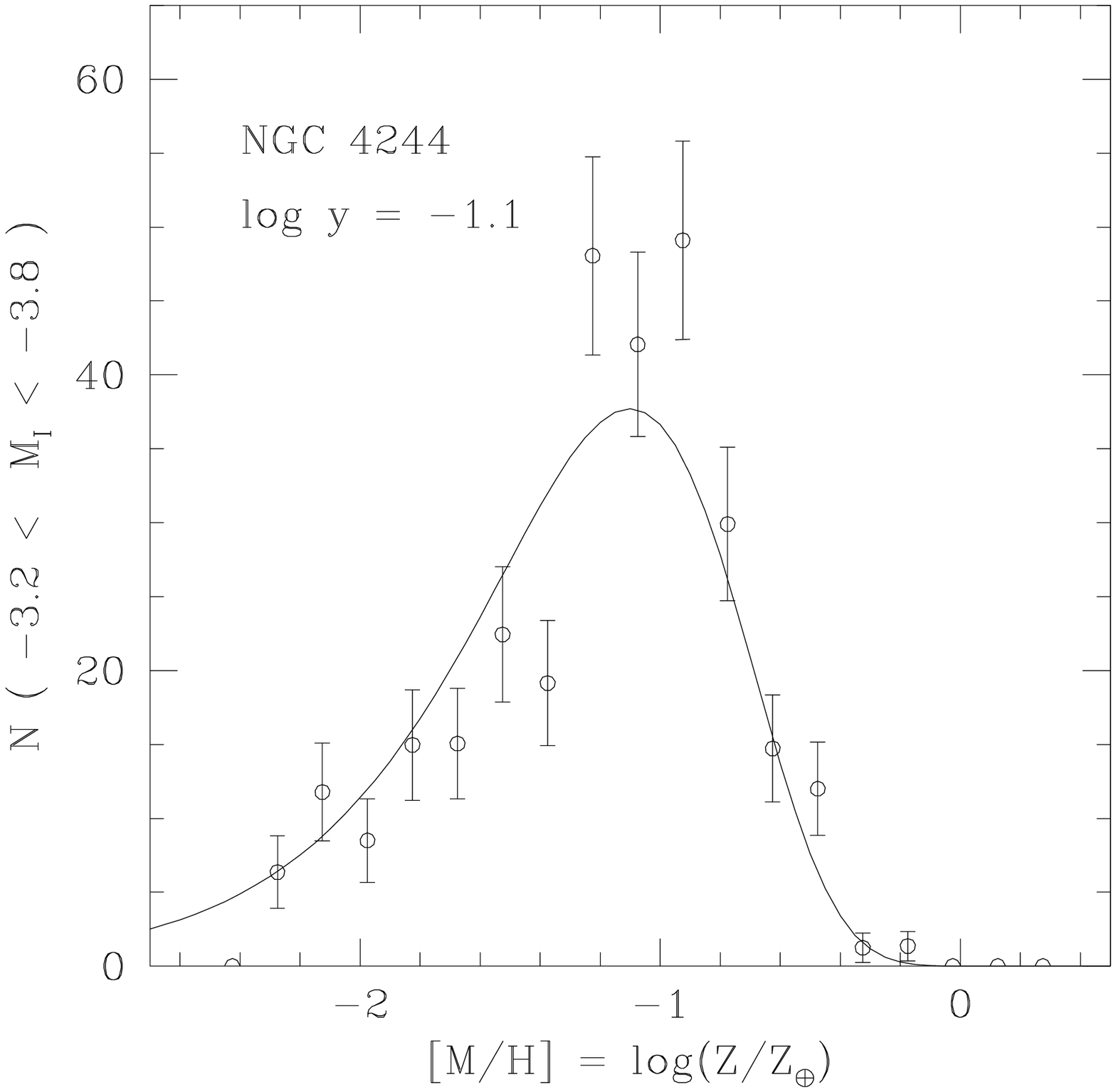}
\includegraphics[height=3.5in]{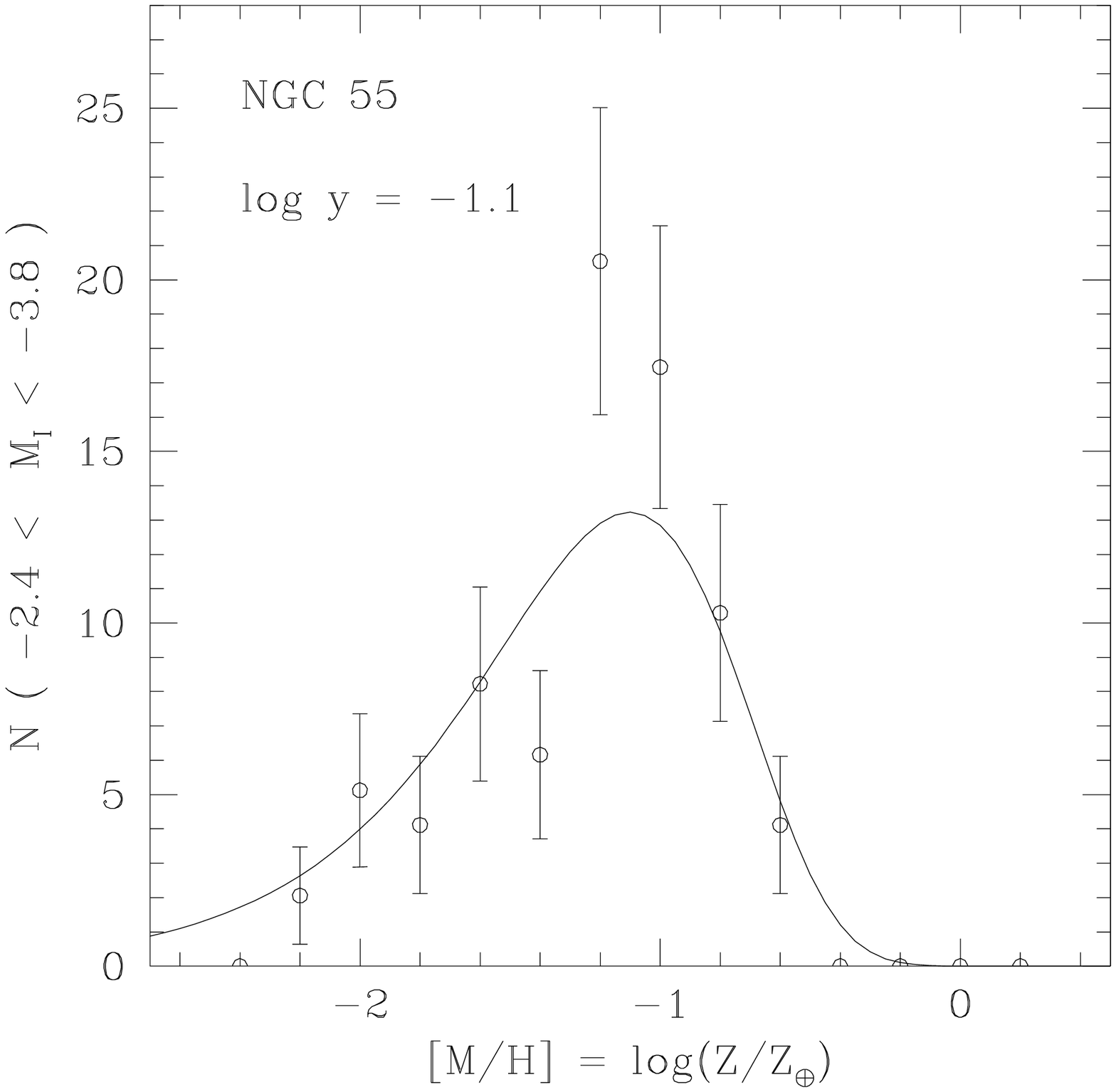}
\includegraphics[height=3.5in]{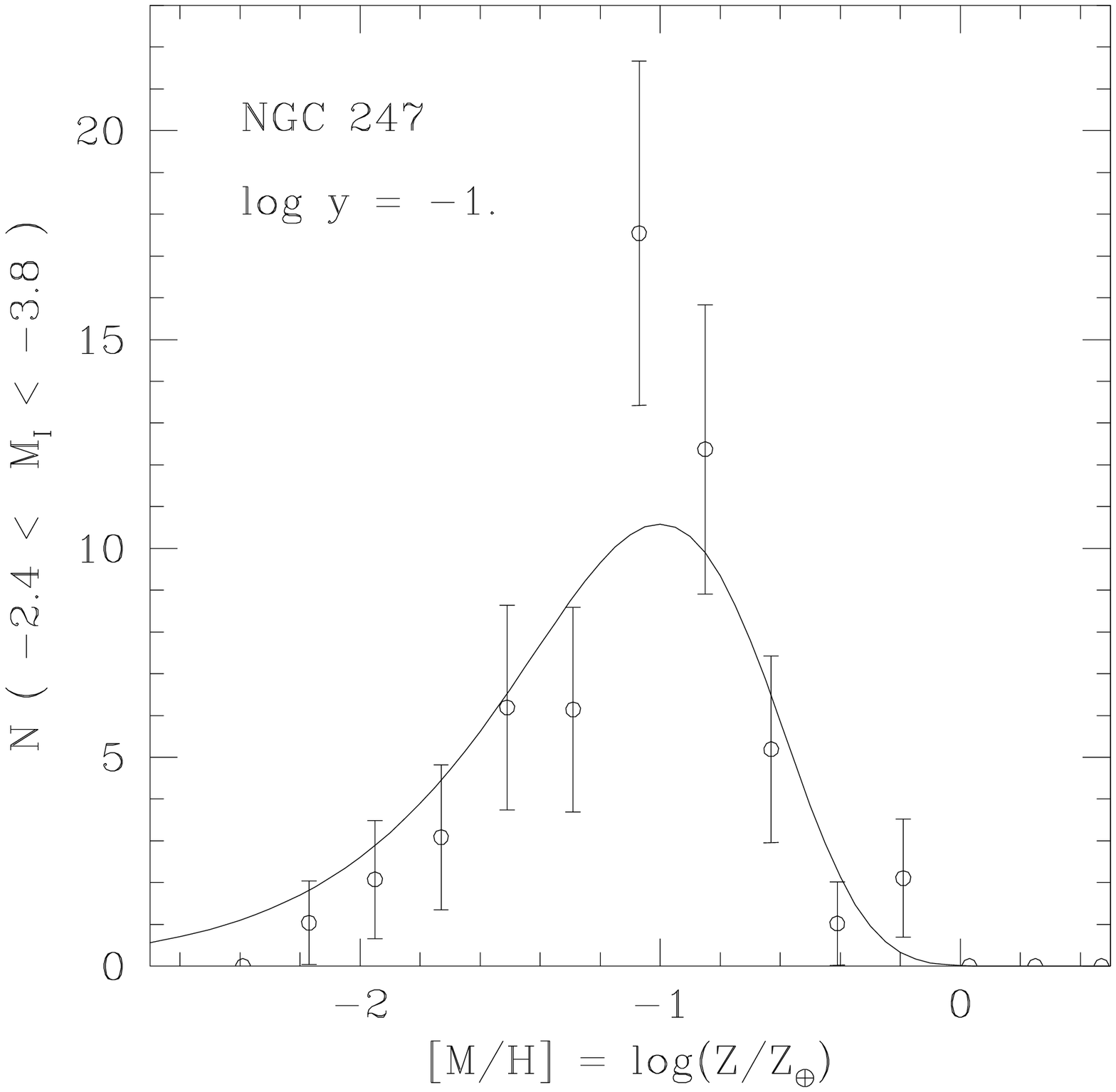}
\includegraphics[height=3.5in]{mdf_ngc300_mdl.ps}
\caption{Same as in Fig.\,\ref{mdf_mdl1} except for NGC~4244, 
NGC~55, NGC~247, and NGC~300.}
\label{mdf_mdl2}
\end{figure}

\begin{figure}
\includegraphics[height=3.5in]{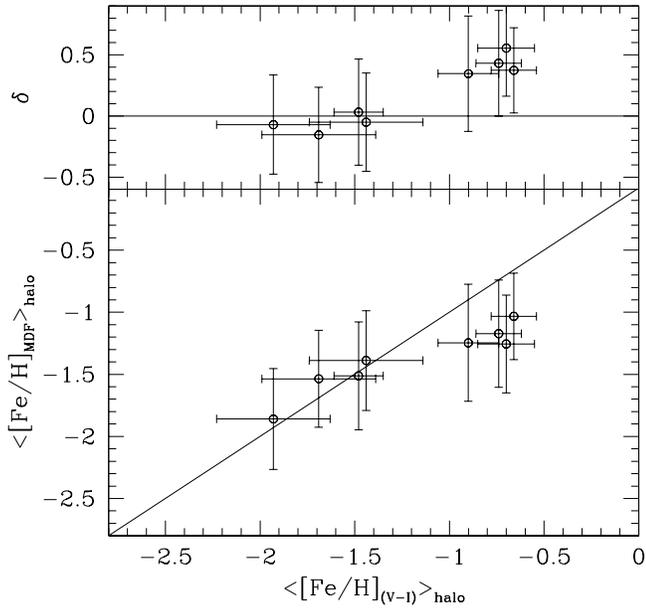}
\caption{Comparison between stellar halo mean metallicities
derived from the colors of the red giant branch stars at a 
luminosity of $M_I = -3.5$, i.e., ${\rm <Fe/H]_{(V-I)}>}$ 
and stellar abundances as calibrated by Lee et al. (1993), 
and these derived from the metallicity distribution functions, 
i.e., ${\rm <Fe/H]_{MDF}>}$.}
\label{comp_feh}
\end{figure}

The resultant incompleteness-corrected metallicity distribution 
functions for all galaxies in our sample are shown in 
Fig.\,\ref{mdf_mdl1} and Fig.\,\ref{mdf_mdl2}, and listed in 
Table\,\ref{gal_mdf}. An optimal distribution bin width was estimated
on a case-by-case basis according to the halo star sample size and 
the distribution's skewness, using the method of Scott (1992).
To account for the effects of incompleteness, we count each star as 
the inverse of the photometric completeness at the location of the 
star on the color-magnitude diagram. Measurement errors are readily 
computed, and $\pm 1 \sigma$ error bars are shown. 
Each panel shows the magnitude range used to estimate the metallicity
distribution function for each galaxy in the sample. The magnitude 
intervals have been chosen to minimize the effects of both the 
photometric errors and the incompleteness.
In the second paper of this series, we have derived the mean 
metallicities of halo stars using the relationship between the mean 
${\rm (V-I)_{\circ}}$ colors of the red giant branch stars at a 
luminosity of $M_I = -3.5$ and stellar abundances as calibrated by 
Lee et al. (1993). The last column of Table\,\ref{gal_prop} 
lists these mean stellar metallicities. The mean metallicities of 
halo stars could also be derived from the metallicity distribution 
functions. In principle, these two mean stellar metallicities should 
be similar. Fig.~\ref{comp_feh} shows a comparison between the two 
mean stellar metallicities. The mean metallicities of halo stars of 
faint galaxies, i.e., ${\rm M_{V,\circ}\sim\,-19}$, agree well with 
the derived mean metallicities of halo stars using the color of red 
giant branch stars at ${\rm M_I = -3.5}$. However, a systematic 
difference at the of 0.2--0.4 dex level is found between the two mean
metallicities of halo stars for bright galaxies, i.e., 
${\rm M_{V,\circ}\sim\,-21}$. The peaks of the metallicity distribution
functions of bright galaxies tend to be more metal-poor than the mean 
metallicities derived using the [Fe/H] vs. ${\rm (V-I)_{\circ,-3.5}}$ 
relationship of Lee et al. (1993).

To calibrate the model grid of VandenBerg et al. (2000) used to derive 
the metallicity distribution functions, Harris \& Harris (2000) have 
compared the well known metallicities of the fiducial Milky Way 
globular clusters to the metallicities derived from ranning their 
red giant branch sequences through the interpolation algorithm (see 
\S\,\ref{calib}). 
At both low- and high-metallicity ends, they found a satisfactory 
agreement between the metallicities derived from the interpolation 
within the VandenBerg et al. (2000) model grid and the literature 
metallicities. However, at intermediate metallicities, i.e., ${\rm 
[Fe/H]\sim\,-0.7}$, the agreement is poorer. The estimated metallicity 
of the Milky Way globular cluster 47~Tuc from the interpolation within 
the model grid is lower than the literature value by $\sim\,0.3$ dex, 
on the order of the observed systematic difference between halo star 
mean metallicities from the metallicity distribution functions and 
those from the red giant branch star colors at a luminosity of 
${\rm M_I=-3.5}$ (that we repored in paper II). 
The mean colors of the observed red giant branchs of bright galaxies 
in our sample agree well with the red giant branch sequence of 47~Tuc 
as shown in Fig.\ref{galcmd1}. Thus, the shift between the two mean 
metallicities of bright galaxy halo stars is due to the fact that 
the VandenBerg et al. (2000) models predict a redder red giant branch 
sequence than the observed one at the 47~Tuc metallicity. 
Thus, if metal rich stars (${\rm [Fe/H]\ga\-0.7}$) are present, the 
mean stellar halo metallicities derived from the colors of red giant 
branch stars at a luminosity of ${\rm M_I=-3.5}$ may be a bit more 
reliable than these derived from the mean of the metallicity 
distribution functions.

Inspecting the metallicity distribution functions, a number of common 
features emerge. The overall shape for all metallicity distribution 
functions is similar; they are characterized by prominent peaks at 
the metal-rich end with extended metal-poor tails. The figures show 
that the position of the dominant peak is a function of the parent 
galaxy luminosity: the brighter the galaxy, the higher the metallicity 
of the prominent peak of the halo metallicity distribution function. 
The I-band magnitude range selected to construct the metallicity
distribution functions of NGC~253 and NGC~4945 stellar halos should 
allow us to detect stars with metallicities up to solar. If super-solar 
stars are present in the halos of these galaxies, we might be missing 
them. The sharp cut-off of the metallicity distribution function at 
the high metallicity end appears to be real, instead of an effect of 
the photometric incompleteness or another observational artifact. 
The evolutionary tracks with metallicities in the range 
$-0.6\le [M/H]\le 0$ are above the magnitude range where the 
incompleteness starts to be severe. 
%
The only galaxy for which the sharp cut-off of the metallicity distribution
function at the high metallicity end could be due to observational effects
is NGC~4258. The metal-rich end of the metallicity distribution function 
of this galaxy is suspected to be incomplete, i.e., a fraction of metal-rich 
stars are missing. For this galaxy, the photometry does not probe the red 
giant branch far, and the 50\% completeness level limit falls close to the 
brightest end of the RGB sequence. We have limited ourselves to a range of
I-band magnitudes where the completeness level is not severely low. However, 
stars with metallicities higher than ${\rm [Fe/H]\sim\,-0.7}$ lie at fainter 
magnitudes than the magnitude range chosen to construct the metallicity 
distribution function of NGC~4258 halo stars. If the halo of this galaxy
contains metal-rich stars, we might be missing a fraction of them. 

%
Similar features are seen for the metallicity distribution functions of both 
M~31 (Durrell et al. 2001, 2004) and NGC~5128 (Harris \& Harris 2000), i.e., 
prominent metal-rich peak, metal-poor tail, and a sharp metal-rich end. 
The metallicity distribution function of the Galaxy stellar halo is distinct 
from those of our galaxy sample and from M~31 and NGC~5128, suggesting again 
that the halo of the Galaxy is not typical for a luminous (spiral) galaxy 
(see also Mouhcine et al. 2004b). 

The most straightforward models of stellar population enrichment considers 
the closed box evolution of a homogeneous proto-galactic gas cloud having 
initial abundance $z_o$. As time advances, the gas is consumed to form stars, 
and more massive stars, that evolve rapidly, return a fraction of their mass, 
enriched in metals due to the nucleosynthesis, to the gas cloud. 
The instantaneous recycling approximation neglects the lifetime of 
stars that contribute to the chemical enrichment compared to any timescale 
of the system, and assumes complete and homogeneous ejecta mixing 
(Searle \& Sargent 1972; Pagel \& Patchett 1975). The simple model 
assumes that no infall and/or outflow of metals has occured during the star 
forming phase of the system. The stellar yield, defined as the ratio of the 
mass of new metals ejected to the mass locked into long-lived stars and 
remnants, parametrizes the model. Under these assumptions, the stellar 
metallicity distribution follows the simple law:

\begin{equation}
f(z)\,=\,(1/y)\exp\left[-(z-z_o)/y\right]
\end{equation}

\noindent
where $y$ is the stellar yield. The yield and the mean abundance $<z>$ are 
related via: $y=<z>-z_{\circ}$, where $z_{\circ}$ is the initial metallicity. 
Hence, the yield is the average metallicity of the stars in a one-zone system 
following the exhaustion of the gas (Hartwick 1976). If metals are lost from 
the system by gas outflow, assuming that the gas loss rate is proportional 
to the star formation rate, the functional form of the metallicity distribution 
is unchanged, but the true yield is reduced to the effective yield 
$y_{eff}=y/(1+c)$, where $c$ is a parameter related to the gas loss rate 
(for a detailed discussion, see Binney \& Merrifield 1998). The only two free 
parameters of the model are the mean metallicity of the stellar population 
and the initial abundance of the protogalactic gas. In the following, we will 
assume that the initial star formation event has consumed a primordial heavy 
element-free gas, i.e., $z_{\circ}=0$. The agreement between the observed 
galaxy luminosity-stellar halo metallicity and the expected scaling from 
Dekel \& Silk (1986), where the chemical evolution of the protogalaxy is 
approximated by this simple model, suggests that it may also provide a 
convenient approximation of the metallicity distribution of halo stars
(see Mouhcine et al. 2004b). 

\begin{figure}
\includegraphics[height=3.5in]{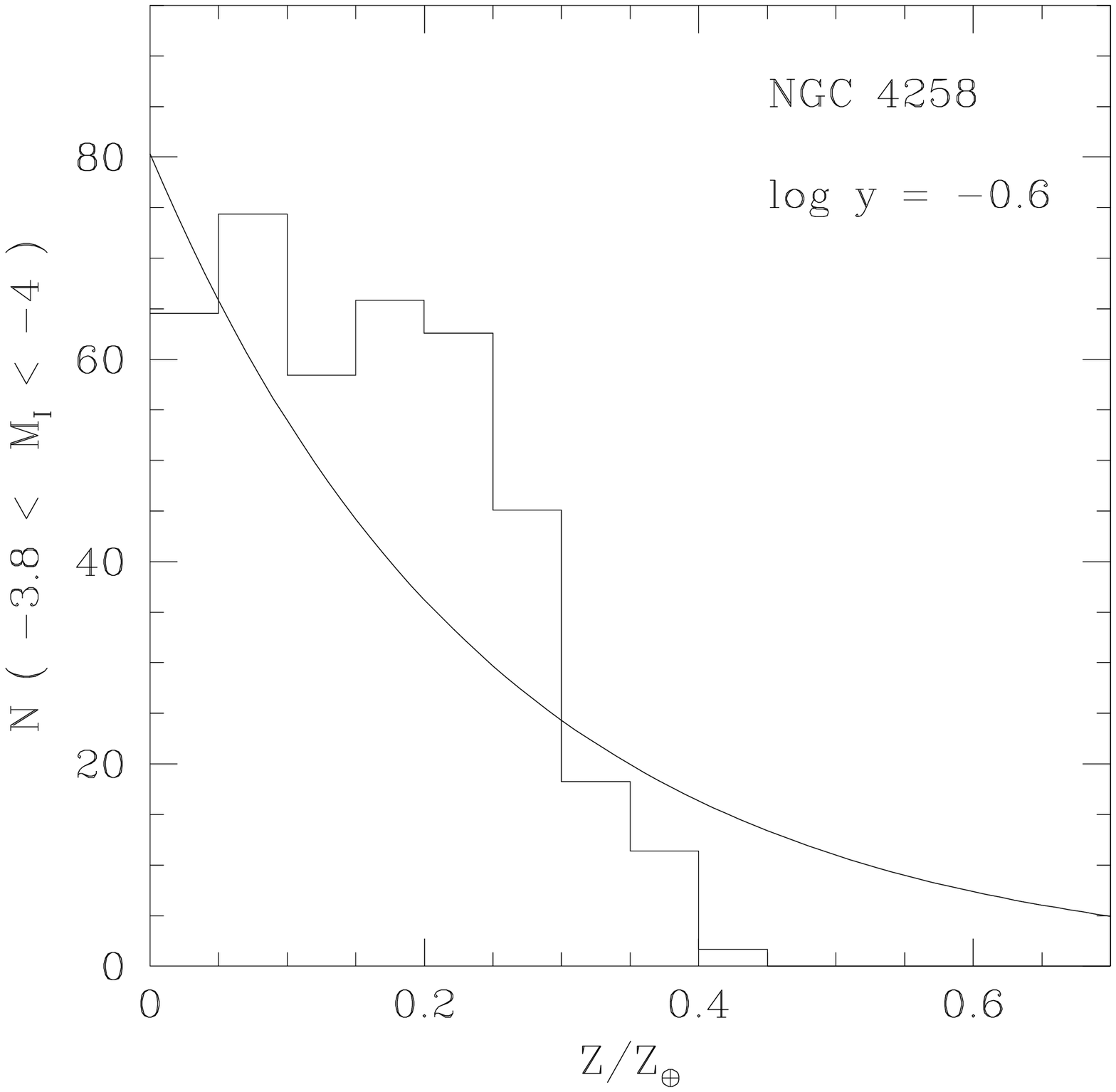}
\includegraphics[height=3.5in]{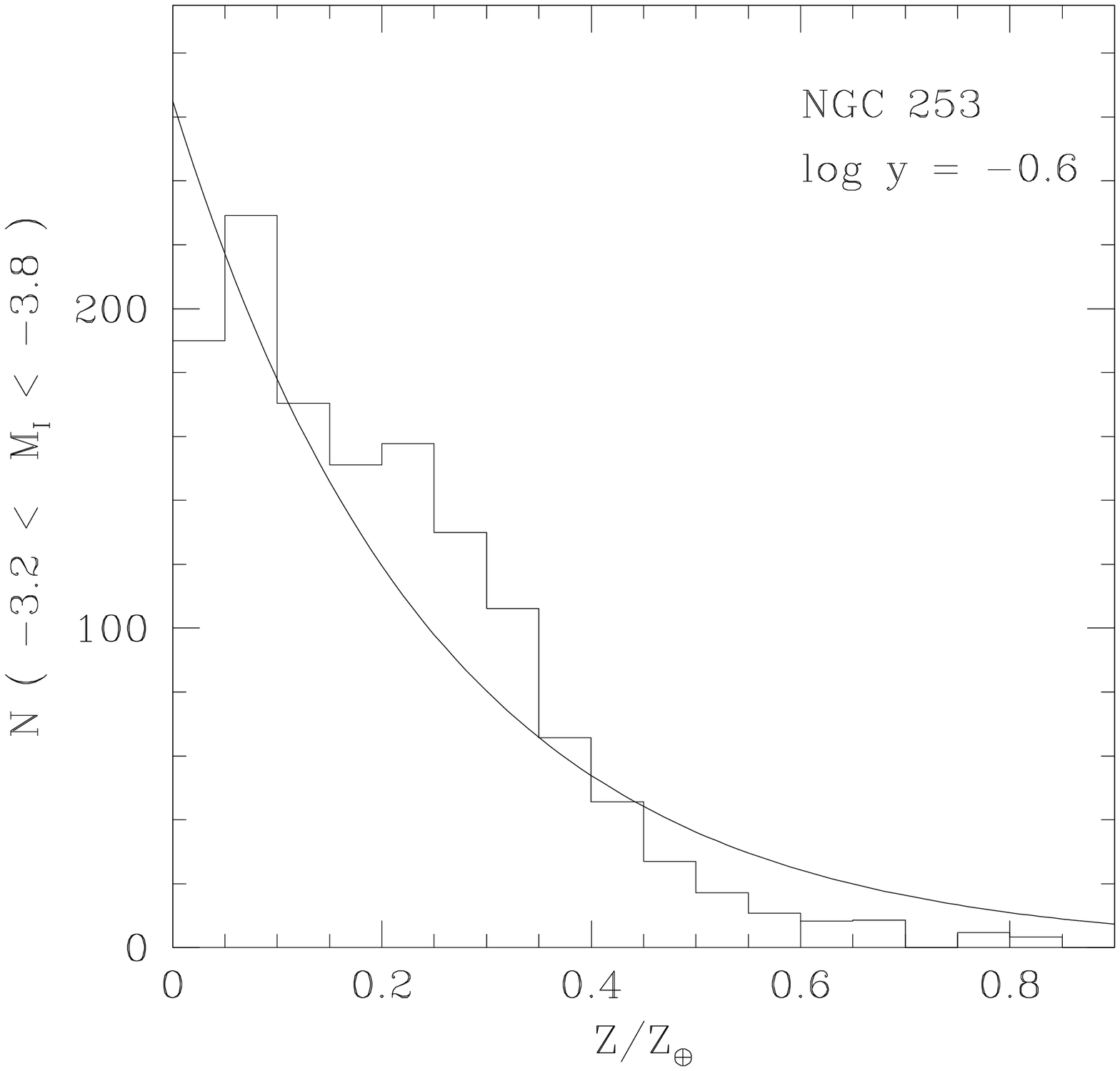}
\includegraphics[height=3.5in]{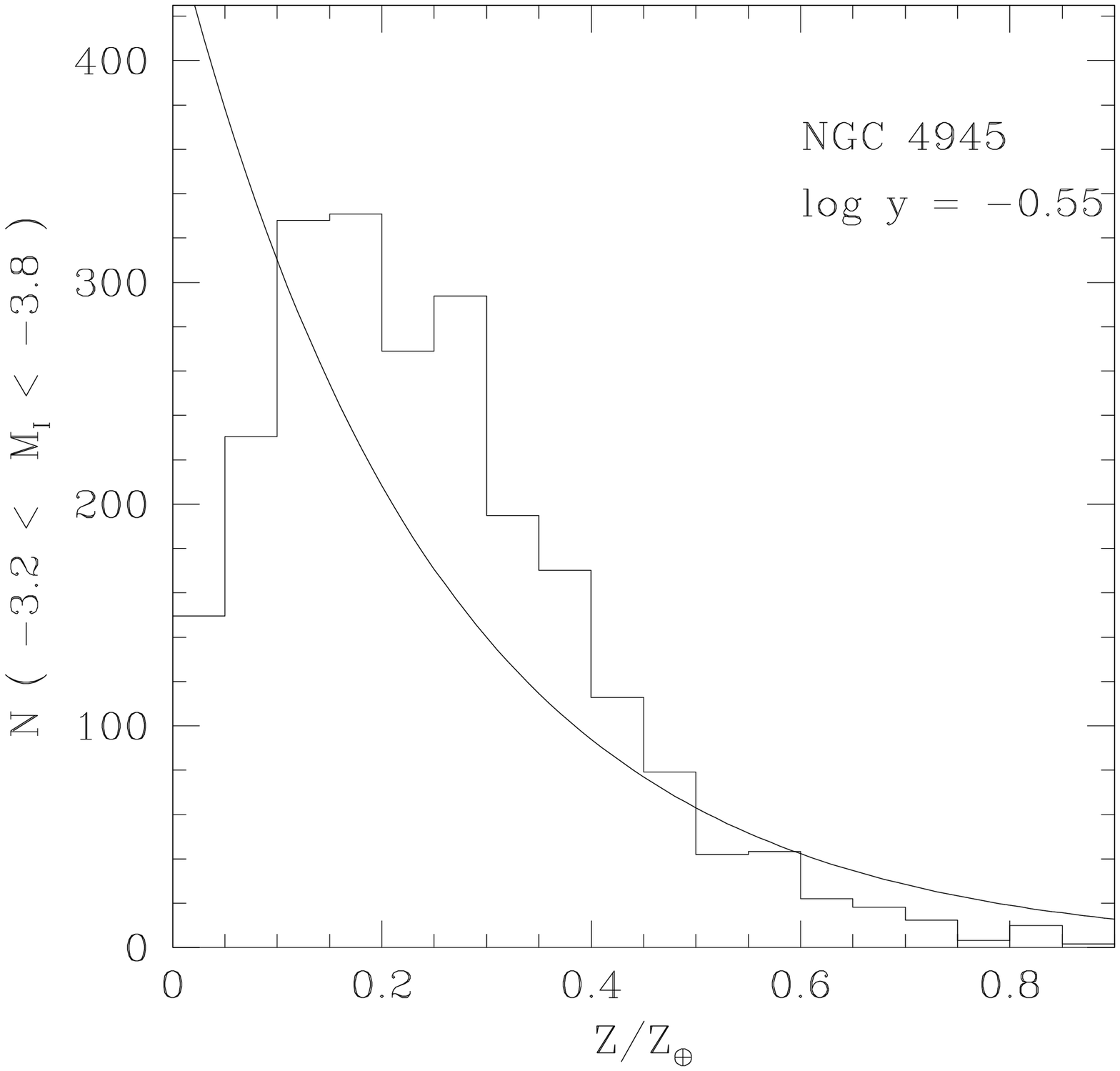}
\includegraphics[height=3.5in]{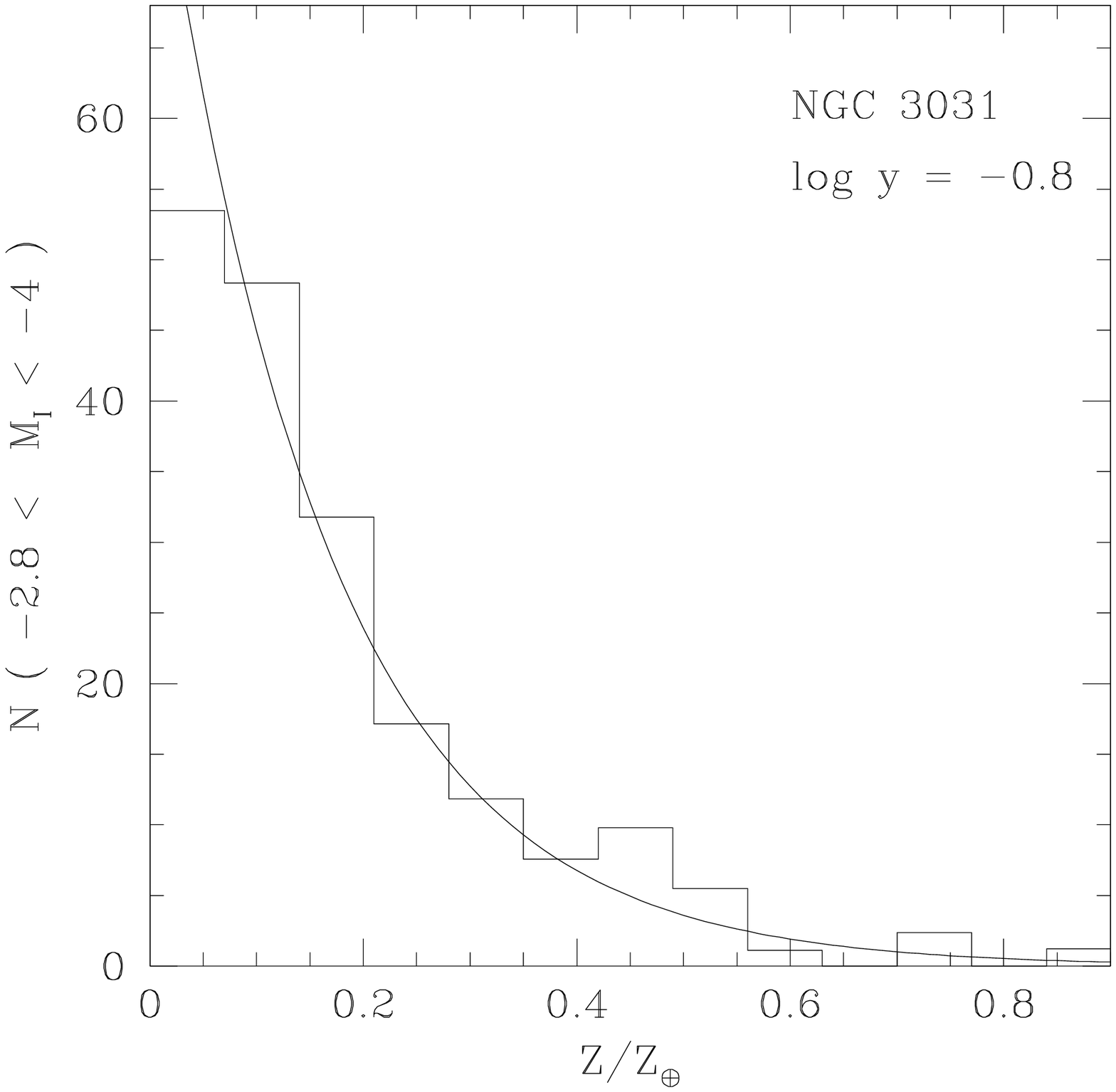}
\caption{Linear metallicity distribution functions for NGC~4258, 
NGC~253, NGC~4945, and NGC~3031. The lines are the prediction 
of the simple chemical evolution model (closed-box model) with 
the indicated value of the yield ($y=<Z>$).}
\label{mdf_lin1}
\end{figure}

\begin{figure}
\includegraphics[height=3.5in]{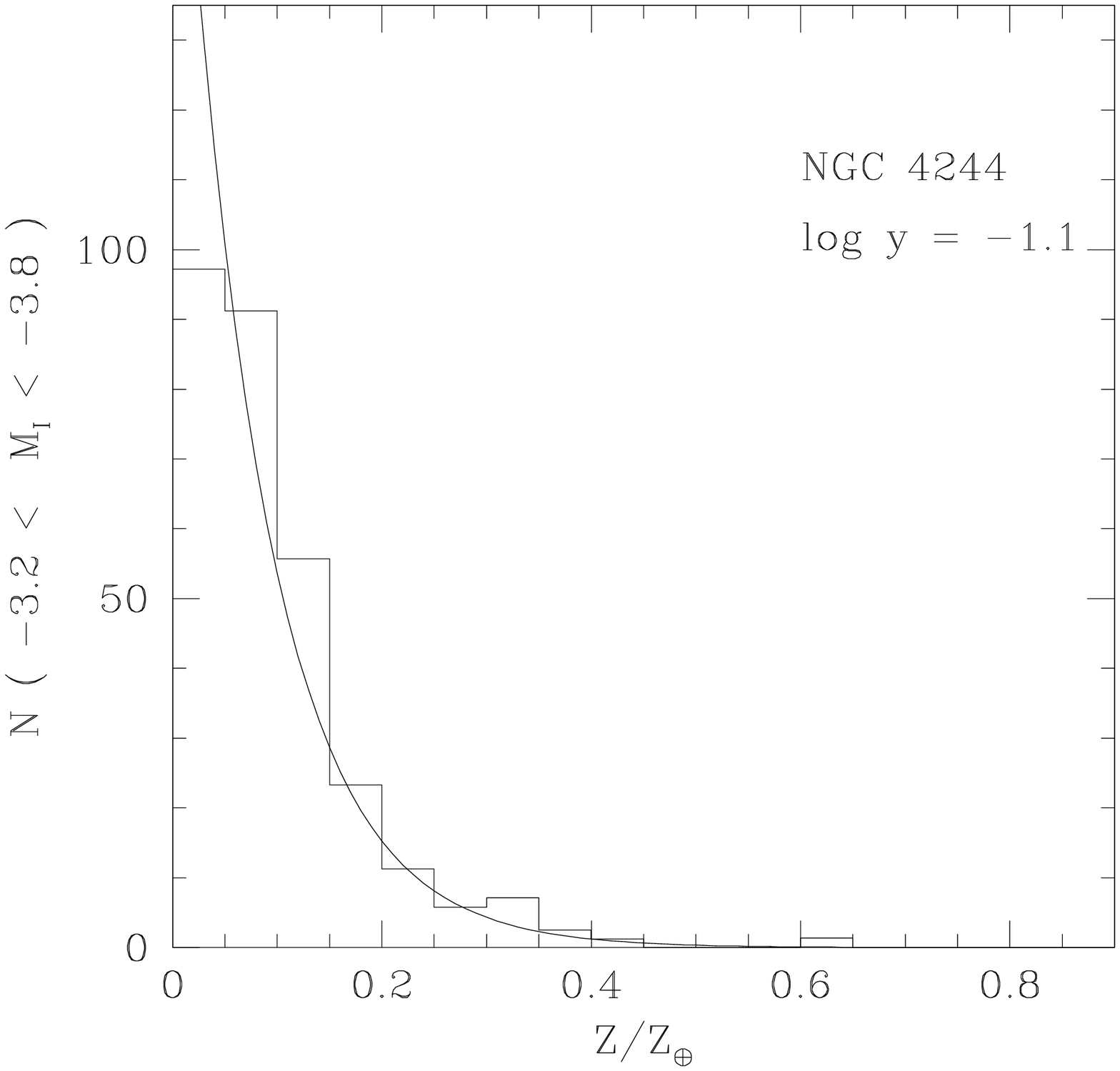}
\includegraphics[height=3.5in]{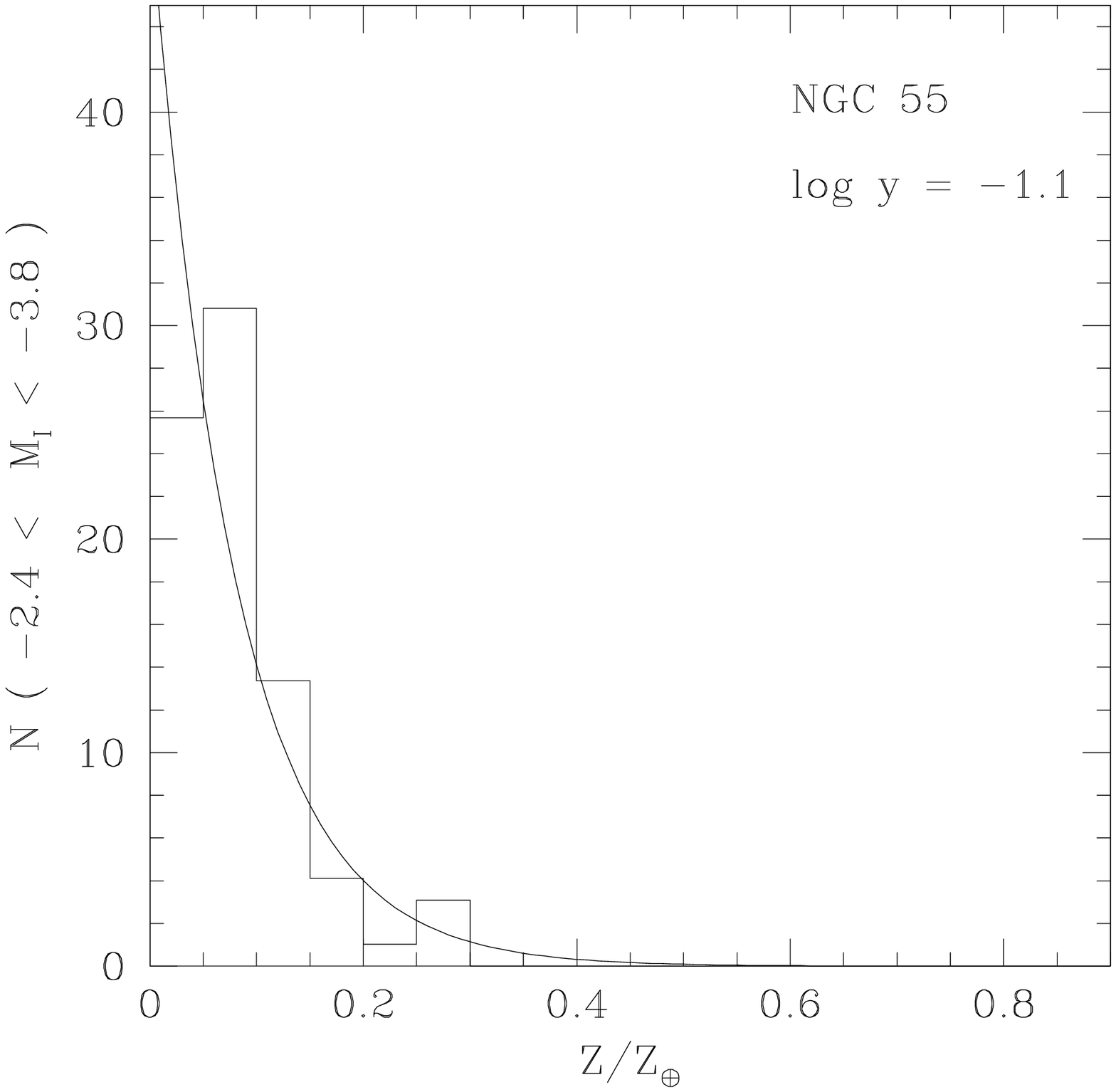}
\includegraphics[height=3.5in]{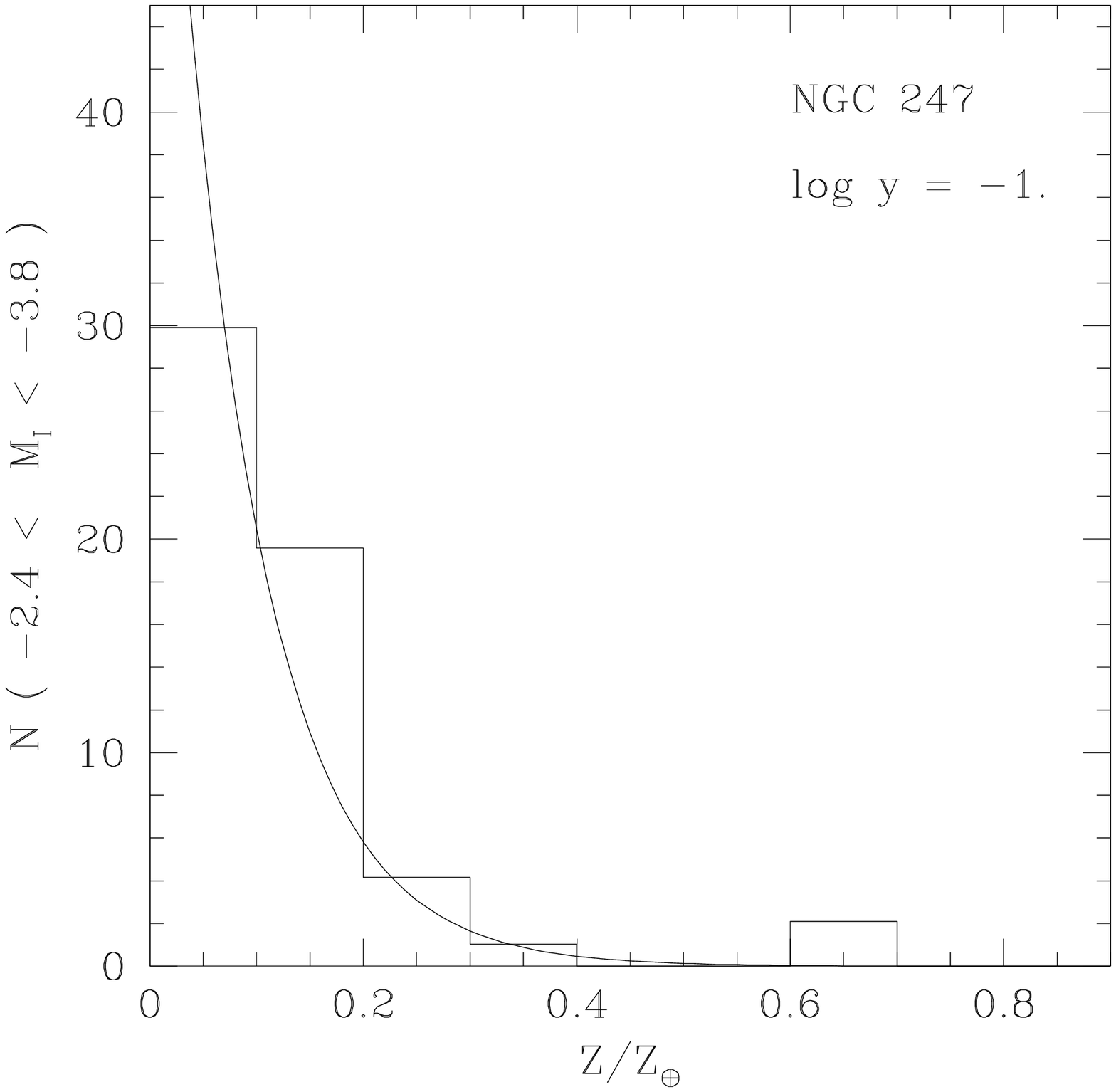}
\includegraphics[height=3.5in]{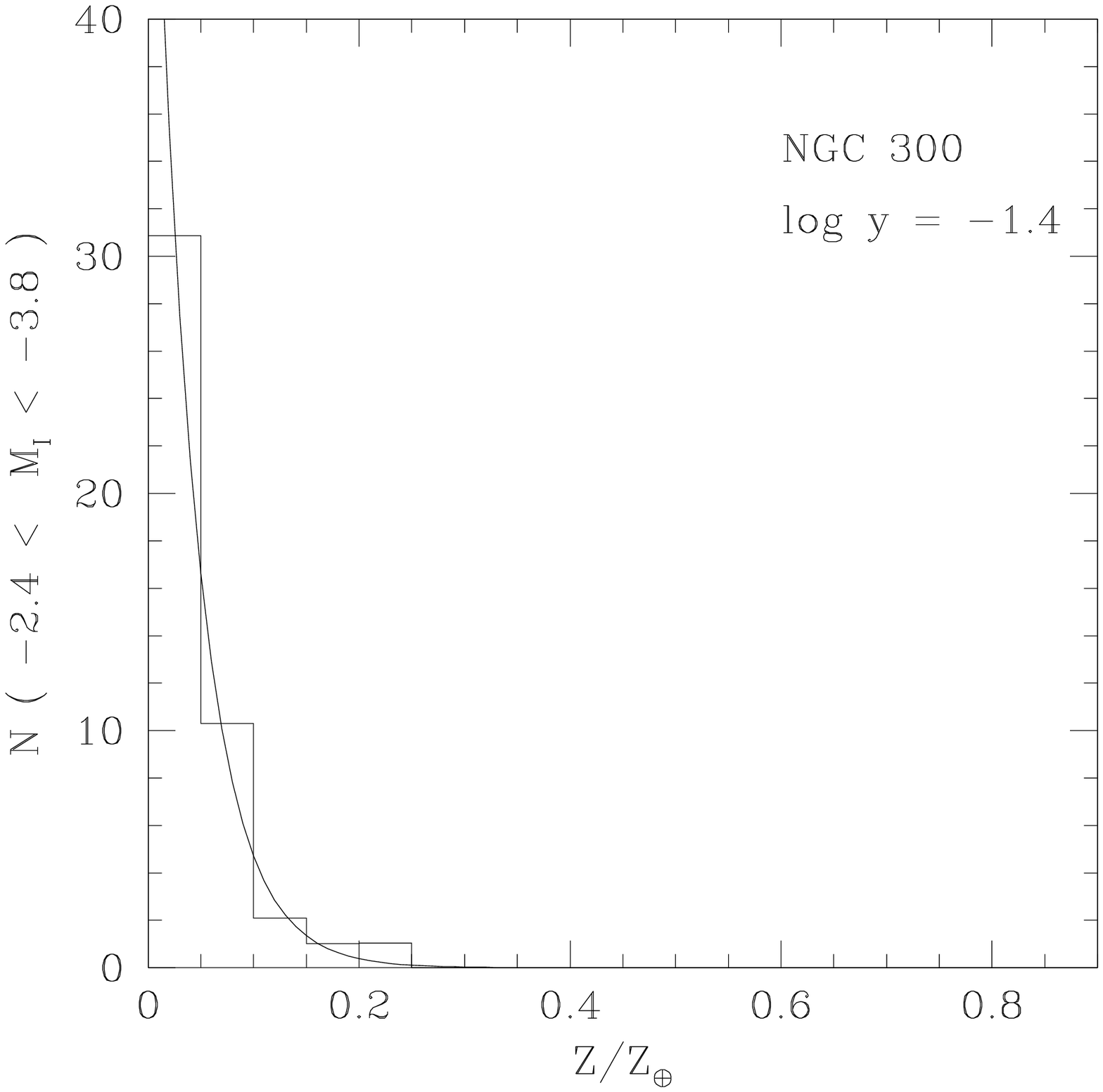}
\caption{Same as in Fig.\,\ref{mdf_lin1} except for 
NGC~4244, NGC~55, NGC~247, and NGC~300.}
\label{mdf_lin2}
\end{figure}

Fig.\,\ref{mdf_mdl1} and Fig.\,\ref{mdf_mdl2} show the simple chemical 
model predictions, with the choice of the effective yield as indicated in 
each panel, overplotted on top of the observed metallicity distribution 
function. The photometric uncertainties for the stars considered here are 
small (see Fig.~\ref{galcmd1} and Fig.~\ref{galcmd2}), and the models have 
not been smoothed.
Fig.\,\ref{mdf_lin1} and Fig.\,\ref{mdf_lin2} show the same models and 
the same data on a linear metallicity scale. As the simple chemical model 
satisfies the requirement of a probability distribution, the model 
predictions are scaled by the area under the data histogram, so that the 
area covered by the data and fits are identical. It is has to be mentioned 
that by comparing the observed metallicity distribution to the simple 
chemical model predictions, we implicitly assume that galaxies are built 
up within a single potential well. 

The figures show that the simple chemical models grossly reproduce 
the general shape of the estimated metallicity distribution function. 
The number of observed stars at the low metallicity end continues to rise 
as the metallicity decreases as required by the model. The effective yields 
needed to reproduce the metallicity distribution functions are a function 
of the parent galaxy luminosity. The effective yield values needed to get 
an agreement with the observed metallicity distribution of bright galaxies 
are similar to those needed to fit the metallicity distribution functions 
of the M~31 halo (Durrell et al. 2001), and NGC~5128 (Harris \& Harris 2000), 
but much larger than the best value for the Galaxy halo. 
   
Despite the success of the simple chemical model to reproduce, in broad 
terms, the metallicity distributions, there are, however, noticeable 
deviations, especially when bright galaxies are considered. The simple 
model predicts that the subsolar regime ($Z < 0.3 Z_{\odot}$) is more 
populated than what is observed, deviating from the exponential-decay 
condition. On the other hand, the number of stars in the intermediate 
metallicity regime ($Z\sim 0.5 Z_{\odot}$), is underestimated by the 
simple chemical evolution models. Quite striking is that the observed 
metallicity distribution is significantly narrower than the simple 
chemical model predictions. At near-solar metallicity, a star excess 
is predicted, and a sharp cut-off is generally missed by the models. 
As discussed above, this sharp cut-off is real; it may be even sharper 
given that any observational bias can only smooth out, and then broaden, 
any sharp feature in the metallicity distribution.   This relative lack
of metal poor stars is evident for the halo of M31, where it is possible
to reach fainter than the horizontal branch.  The metal poor fraction,
as evidenced by the fraction of blue HB stars, is $<15\%$ 
(Bellazini et al. 2003)

\section{Interpretation: implication for galaxy formation}

What kind of stellar halo formation scenario can account for the new 
observations? A key element to understanding the formation of stellar halos 
can be found in the shape of their MDFs. The stellar halos of bright 
galaxies are dominated by a metal-rich stellar population. This implies 
that those galaxies underwent a large amount of chemical enrichement before 
the stellar halos were assembled, or that the halos were assembled from
metal-rich satellites. Hierarchical models of galaxy formation 
provide a reasonable picture of galaxy evolution. C\^ot\'e et al. 
(2000) have discussed a model for the formation of a galactic spheroid, 
where it is assembled by dissipationless accretion of isolated, chemically 
distinct, small and metal-poor protogalactic fragments. These models are 
able to reproduce the metallicity distributions of both diffuse halo stars
and halo globular cluster systems in both the Galaxy and M~31 provided 
that the luminosity function of protogalactic fragments that were accreted 
to form the halo of M~31 is flatter than what is needed to reproduce the 
Galaxy halo metallicity distributions. 
Despite its success in accounting for the overall shape of the MDFs of M~31 
and Milky Way stellar halos, this model has
difficulties accounting for the large excess 
of stars at $Z\approx 0.3-0.6 Z_{\odot}$, i.e., the peak of the metallicity 
distribution (see also Durrell et al. 2001). These stars are too metal-rich 
to be formed in a shallow potential well, i.e., in a low-mass protogalactic 
fragment. The comparison between the observed stellar halo MDFs and those 
of dwarf galaxies, i.e., NGC~147 and BK~5N in the M~81 group as given in 
Harris \& Harris (2000), show clearly that stellar halos contain a rather 
small number of metal-poor stars that dominate dwarf galaxies. 
This suggests that the stellar halos of spiral galaxies, at least the bright 
ones, have not been predominantly assembled from the disruption of 
metal-poor dwarf satellite galaxies similar to ones observed in the Local 
Group and M~81 galaxy group. The differences of chemical abundance ratios 
between Galaxy halo stars and those of Local Group dwarf spheroidal galaxy 
stars seems to indicate that this may be also the case for the Milky Way 
stellar halo (Shetrone et al. 2003; Tolstoy et al. 2003). 
Harris \& Harris (2001) have shown that NGC~5128 stellar halo MDF is 
statistically indistinguishable from the MDF of the Large Magellanic Cloud 
(LMC) outer disk stars constructed using a sample of giant stars from Cole 
et al.(2000). The observed metallicity distributions of bright galaxy stellar 
halos  share similarities with the metallicity distributions of both the inner 
region of the faint compact elliptical galaxy M~32 (Grillmair et al. 1996) and 
the LMC outer disk stars (Cole et al. 2000), i.e., small number of metal-poor 
stars, and the predominance of ${\rm [M/H]\sim\,-0.5}$ stars. These similarities 
suggest that at least a fraction of halo stars originates from the disruption 
of LMC-like galaxies. This is in agreement with the suggestion of Ibata et al. 
(2001) and Harris \& Harris (2001) that the bulk of the halos of M~31 and 
NGC~5128 may have formed from a single event, such as the disruption of 
similar metal-rich stellar systems.  
Yet another possibility is that the stars formed from gas which was
pre-enriched by the formation of the spheroid.  The high metallicity is 
suggestive of gas originating in violent starburst that may be connected
with the formaton of the bulge.

The sharp high-metallicity cutoff of the metallicity distribution, in 
comparison to the predictions of simple chemical evolution model, suggests 
that the star formation did not proceed to exhaust entirely the gas reservoir. 
If spiral stellar halos formed in a rapid and violent star-forming event, 
one may expect that a fraction of the heavy elements, produced by supernovae, 
are lost by means of metal-rich galactic winds before the metal-rich gas was 
exhausted by star formation. 
Another possible way to produce the apparent pronounced peak in the 
metallicity distribution functions may be the presence of a large age spread 
for halo stellar populations. If metal-rich halo populations are significantly 
younger than halo metal-poor populations, metal-rich RGB stars will be bluer 
than they would be if they were old. Thus, their metallicity will be 
systematically underestimated if the metallicity distribution functions are 
constructed following the assumption that halo field stars are uniformely old. 
Deep observations of the M~31 stellar halo indicate the presence of an age 
spread larger than few Gyr (Brown et al. 2003). Unfortunately, no constraints 
are available on the age distributions of halo stellar populations for other 
galaxies. Deep and wide field observations of stellar halos are needed to 
get insight into the age distribution of halo stellar populations over a 
significant fraction of the halo.  

The paucity of metal-poor stars compared to the simple chemical model 
predictions may flag the presence of a G-dwarf like problem, similar to 
what is observed for the Galactic disk (Tinsley 1980), and also possibly
in the bulge (Zoccali et al. 2003). Investigations of integrated light 
properties show that the G-problem is present in early type galaxies 
(Greggio 1997).
It is intriguing to see that stellar systems with different histories are 
sharing the same problem, the reason for which is not fully understood. 
For galactic stellar halos, pre-enrichment of the gas from which halo 
stars form can be invoked to explain the moderate number of metal-poor 
stars compared to the simple chemical model predictions. The closed-box 
chemical evolution model stands on the assumption that the gas mass is 
assembled entirely before the star formation starts, i.e., no outflow 
or inflow. It is well established that the G-dwarf problem is eliminated 
if the baryonic matter assembly is assued to be gradual (Pagel 1997).  
This suggests that stars which end up in the halo of spiral 
galaxies were formed gradually within their formation site(s).

\section{Summary \& Conclusions}
\label{concl}

We have presented the results from a study of the stellar populations within 
the halos of spiral galaxies. The main goal of our study is to investigate 
the properties of spiral halos, and to deepen our understanding of the 
physical processes that govern their formation, and to shed light on the 
buildup of galaxies over the cosmic time. 

We have used WFPC2 to image in the F606W ($V$) and F814W ($I$) filters halo 
fields in a sample of eight nearby highly inclined galaxies, to construct 
the color-magnitude diagram for the halo diffuse stellar populations in 
spiral galaxies. The observations were designed in a way that the resultant
color-magnitude diagrams are complete to at least a magnitude below the tip 
of the red giant branch. 
Such depth secures the usage of the red giant branch stars to investigate 
the properties of stellar halos. With this galaxy sample and data quality, 
it is possible to begin addressing systematically questions regarding the 
correlations between the properties of spiral halo field stellar populations 
and spiral galaxy properties. 

We have found that the halo color-magnitude diagrams are morphologically 
dominated by old red giant branch stars, with no significant number of 
halo stars brighter than the classic old tip of the red giant branch. 
This is consistent with the conventional interpretation of galaxy halos 
being composed primarily by an homogeneous, old stellar population. 
However, the color range spanned by giant stars is sensitive to the parent 
galaxy luminosity. The stellar halos of bright galaxies contain red giant 
branch stars with redder colors than faint galaxies.

A finely spaced grid of theoretical evolutionary tracks together with 
an empirical red giant branch of a Galactic metal-rich cluster of a known 
metallicity, was used to convert the color-magnitude diagrams of the halo 
stellar populations into photometric metallicity distributions. 
The distributions of stellar metallicities are found to show a broad peak 
at the metal-rich end, with low metallicity tails extending to 
${\rm [Fe/H]\sim\,-2}$, with a fairly sharp edge at the high-metallicity 
end. The metallicity of the metal-rich peaks depend on the parent galaxy 
luminosity. The halo stellar populations are dominated by moderately 
high-metallicites, i.e., ${\rm -0.6<[M/H]<0}$. This suggests that stellar 
halos are more likely to be formed by the disruption of intermediate mass 
galaxies, i.e., similar to M~32 and LMC, rather than the accretion of 
metal-poor dwarf satellite galaxies.
The simple chemical model does match the metallicity distribution to first 
order, with the effective yields best matching the observed distributions 
increasing with the galaxy luminosity. The analysis of the departures of
observed metallicity distributions from the predictions of the simple 
chemical models indicates that the halo may have built up gradually from 
a pre-enriched gas.   

A high priority for future work should be the extension of this type
of analysis to other spiral galaxies, targeting a variety of
morphological types, disc and bulge properties, and position on the
Hubble sequence.  This will more fully reveal the variation and trends
in halo properties, and to what extent the halo of the Milky Way is unique.

\acknowledgments

M.M would like to thank Rodrigo Ibata for useful and enlightening
discussions. We acknowledge grants under HST-GO-9086 awarded by the Space 
Telescope Science Institute, which is operated by the Association of the 
Universities for Research in Astronomy, Inc., for NASA under contract 
NAS 5-26555

\clearpage

\begin{deluxetable}{lcccccccc}
\tablewidth{6.8in}
\tablecaption{Properties of the sample galaxies. Columns: (1) Galaxy 
name; (2) Morphological type; (3) inclination angle; (4) distance modulus; 
(5) foreground extinction (Schlegel et al. 1998); (6) Galactic fourground 
extinction-corrected V-band absolute magnitude; (7) mean stellar halo
metallicity estimated from the $(V-I)_{\circ}$ color of red giant branch 
stars at a luminosity of $M_I = -3.5$ (see paper II of this series for 
more details); (8) mean stellar halo metallicity estimated as the median of
the metallicity distribution function. }
\tablehead{
\colhead{Galaxy}             & 
\colhead{Type}               & 
\colhead{$i$}                & 
\colhead{$(m-M)_{\circ}$}    & 
\colhead{$E(B-V)^{a}$}       & 
\colhead{$M_{VO}^{b}$}       &
\colhead{${\rm <[Fe/H]>_{(V-I)_{\circ,-3.5}}}$ } & 
\colhead{${\rm <[Fe/H]>_{MDF}} $} }

\tablecolumns{8}
\startdata
NGC 55    &  Sc          & 90 & 26.11 & 0.01   & -19.02  & -1.69 & -1.54 \\ 
NGC 247   &  Sc(s)III-IV & 80 & 27.30 & 0.02   & -18.91  & -1.44 & -1.39 \\ 
NGC 253   &  Sc(s)       & 86 & 27.59 & 0.02   & -21.13  & -0.74 & -1.17 \\ 
NGC 300   &  Sc(s)II     & 46 & 26.53 & 0.014  & -18.62  & -1.93 & -1.86 \\ 
NGC 3031  &  Sb(r)I-II   & 60 & 27.80 & 0.08   & -21.14  & -0.90 & -1.25 \\ 
NGC 4244  &  ScdIII      & 90 & 27.88 & 0.02   & -18.96  & -1.48 & -1.51 \\ 
NGC 4258  &  Sb(s)II     & 71 & 29.32 & 0.02   & -21.30  & -0.70 & -1.26 \\ 
NGC 4945  &  Sc(s)I-II   & 90 & 27.56 & 0.18   & -20.77  & -0.66 & -1.03 \\ 

\enddata
 
\label{gal_prop}
\end{deluxetable}

\begin{deluxetable}{lccccccccc}
\tablewidth{6.5in}
\tablecaption{Incompleteness-corrected metallicity distribution functions 
for the sample of galaxies.  }
\tablehead{
\colhead{[M/H]}    & 
\colhead{NGC 253}    & 
\colhead{NGC 3031}   & 
\colhead{NGC 4258}   & 
\colhead{NGC 4945}   & 
\colhead{NGC 4244}  & 
\colhead{NGC 300}  &
\colhead{NGC 247}  &
\colhead{NGC 55}  }
\tablecolumns{4}
\startdata
-2.45  &     &      &  0.0 & 0.0 & & & & \\
-2.445 &     &      &      &     &  0.0 & & & \\
-2.425 &  0.0 & 0.0 &      &     &      &  & & \\
-2.35  &     &      &  1.1 & 0.0 & & & & \\
-2.295 &     &      &      &     &  6.4 & & & \\
-2.275 &  5.7 & 2.1 &      &     &  &  &  & \\ 
-2.25  &     &      &  1.2 & 3.5 & & & & \\
-2.2 & & & & & & & & 2.1\\
-2.17 & & & & & & & 1.0 & \\
-2.15  &     &      &  2.3 & 5.8 & & & & \\
-2.145 &     &      &      &     &  11.8 & & & \\
-2.125 & 13.9 & 5.1 &      &     &      & 8.24 &  \\ 
-2.05  &     &      &  5.7 & 10.4 & & & & \\ 
-2.0 & & & & & & & & 5.1\\
-1.995 &     &      &      &     &  8.5& & & \\
-1.975 & 20.9& 5.3  &      &     &      &  & &  \\ 
-1.95  &     &      &  1.2 & 8.2 & & & 2.1& \\
-1.875  & & & & & & 10.3& & \\
-1.85  &     &      &  2.3 & 8.1 & & & & \\
-1.845 &     &      &      &     &  15.0 & & & \\
-1.825 & 27.7& 3.1  &      &     &      &  &  & \\ 
-1.80 & & & & & & & & 4.1\\
-1.75  &     &      &  9.1 & 15.0 & & & & \\
-1.73  & & & & & & & 3.1 & \\
-1.695 &     &      &      &     &  15.1 & & & \\
-1.675 & 42.9 & 8.4 &      &     &  &  & & \\ 
-1.65  &     &      &  8.0 & 10.5 & & & & \\ 
-1.625 & & & & & & 6.2& & \\
-1.6 & & & & & & & & 8.2\\
-1.55  &     &      &  4.6 & 17.3 & & & & \\
-1.545 &     &      &      &     &  22.4 & & & \\
-1.525 & 39.6 & 3.1 &      &     &  &  & & \\
-1.51  & & & & & & & 6.2 & \\
-1.45  &     &      & 11.5 & 25.4 & & & & \\
-1.4 & & & & & & & & 6.2\\
-1.395 &     &      &      &     &  19.2 & & & \\
-1.375 & 39.3&10.5  &      &     &  & 8.2 &  &  \\ 
-1.35  &     &      &  17.4& 45.3 & & &  & \\
-1.29 & & & & & & & 6.1 & \\
-1.25  &     &      &  24.5& 48.6 & & & &  \\
-1.245 &     &      &      &     &  48.1 & & & \\
-1.225 & 93.3&16.8  &      &     &     &  & & \\ 
-1.2 & & & & & & & & 20.5\\
-1.15  &     &      &  23.7& 80.4 & & & & \\ 
-1.095 &     &      &      &     &  42.1  & & & \\ 
-1.075 &135.9 &19.9 &      &     &  &  & & \\ 
-1.07 & & & & & & & 17.5 & \\
-1.05  &     &      &  26.1& 101.4 & & & & \\
-1.00 & & & & & & & & 17.5\\
-0.95  &     &      &  26.6& 176.5 & & & & \\
-0.945 &     &      &      &     &  49.1 & & & \\
-0.925 &142.2 &27.4 &      &     &  &  &  & \\
-0.875 & & & & & & 3.1& &  \\  
-0.85  &     &      &  38.1& 202.7 & & & 12.4& \\
-0.80 & & & & & & & & 10.3\\
-0.795 &     &      &      &     &  29.1 & & & \\
-0.775 &179.3 &27.5 &      &     &     &  & & \\ 
-0.75  &     &      &  57.0& 278.5 & & & & \\ 
-0.65  &     &      &  67.9& 281.3 & & & & \\
-0.645 &     &      &      &     &  14.7 & & & \\
-0.63  & & & & & & 5.2& &  \\  
-0.625 &242.1 &21.4 &      &     &      & 1.0 & & \\ 
-0.60 & & & & & & & & 4.1\\
-0.55  &     &      &  51.3& 358.9 & & & & \\
-0.495 &     &      &      &     &  12.0 & & & \\
-0.475 &217.5& 17.2 &      &  &  &  & &  \\ 
-0.45  &     &      &  20.7& 286.4 & & & & \\
-0.41 & & & & & & 1.0& &  \\  
-0.35  &     &      &  1.7 & 194.6 & & & & \\
-0.345 &     &      &      &      &  1.2 & & & \\
-0.325 & 93.8 &17.5 &      &      &     &  &  & \\ 
-0.25  &     &      &  0.0 & 97.2 & & & & \\ 
-0.195 &     &      &      &      &  1.4 & & & \\
-0.19 & & & & & & & 2.1& \\
-0.175 & 28.5 & 3.5 &      &  &  &  &  & \\ 
-0.15  &     &      &  0.0 & 44.2 & & & & \\ 
-0.05  &     &      &  0.0 & 11.6 & & & & \\
-0.025 &  3.4& 1.2  &       &     &    &  &  & \\ 
\enddata
 
\label{gal_mdf}
\end{deluxetable}

\clearpage

\begin{thebibliography}{}
\bibitem[xxx]{xxx} Beers, T.C, \& Sommer-Larsen, J., 1995, \apjs, 96, 175
\bibitem[xxx]{xxx} Bekki, K., Chiba, M., 2001, \apj, 558, 666 
\bibitem[xxx]{xxx} Bellazzini, M., Cacciari, C., Federici, L., FuciPecci, F., 
                   Rich, M.R., 2003, \aap, 405, 867
\bibitem[xxx]{xxx} Bergbusch, P.A., \& VandenBerg D.A., 2001, \apj, 556, 322
\bibitem[xxx]{xxx} Binney, J., \& Merrifiled, M., 1998, Galactic Astronomy 
                   (Princeton: Princeton Univ. Press)	   
\bibitem[xxx]{xxx} Brown, T.M., Ferguson, H.C., Smith E., et al., 2003, \apj, 
                   592, L17
\bibitem[xxx]{xxx} Carney, B.W, Laird, J.B., Latham, D.W., Aguilar, L.A., 
                   1996, \aj, 112, 668
\bibitem[xxx]{xxx} Casertano, S., et al., 2000, 2000, \aj, 120, 2747
\bibitem[xxx]{xxx} Chaboyer, B., Green, E.M., Liebert, J., 1999, \aj, 117, 1360   
\bibitem[xxx]{xxx} Chiba, M., \& Beers, T.C., 2000, \aj, 119, 2843
\bibitem[xxx]{xxx} Caldwell, N., Armandroff, T.E., Da Costa, G.S., Seitzer, P., 
                   1998, \aj, 115, 535
\bibitem[xxx]{xxx} Cohen, J.G., Gratton, R.G., Behr, B.B., \& Carretta, E.,
                   1999, \apj, 523, 739
\bibitem[xxx]{xxx} Cole, A.A., Smecker-Hane, T.A., Gallagher, J.S., 2000, 
                   \aj, 120, 1808   
\bibitem[xxx]{xxx} C\^ot\'e, P., Marzke, R.O., West, M.J., Minniti, D., 2000, 
                   \aj, 533, 869 
\bibitem[xxx]{xxx} Da Costa, G.S., \& Armandroff, T.E., 1990, \aj, 100, 162   
\bibitem[xxx]{xxx} Davidge, T.J., \& Courteau, S., 2002, \aj, 123, 1438
\bibitem[xxx]{xxx} Dekel, A., \& Silk, J., 1986, \apj, 303, 39
\bibitem[xxx]{xxx} Durrell, P.R., Harris, W.E., \& Pritchet, C.J., 2001, \aj, 
                   121, 2557
\bibitem[xxx]{xxx} Durrell, P.R., Harris, W.E., \& Pritchet, C.J., 2004, \aj, 
                   128, 260  
\bibitem[xxx]{xxx} Eggen, O., Lynden-Bell, D., \& Sandage A., 1962, \apj, 
                   136, 748
\bibitem[xxx]{xxx} Fagotto, F., Bressan, A., Bertelli, G., \& Chiosi, C., 1994,
                   \aaps, 104, 365		   
\bibitem[xxx]{xxx} Ferguson, A.M.N., Irwin M.J., Ibata, R.A., Lewis, G., 
                   \& Tanvir, N., 2002, \aj, 124, 1452
\bibitem[xxx]{xxx} Ferrarese, L., et al., 2000, \apjs, 128, 431	   
\bibitem[xxx]{xxx} Fleming, D.E.B., Harris, W.E., Pritchet, C.J., \& Hanes, D.A.,
                   1995, \aj, 109, 1044 	   
\bibitem[xxx]{xxx} Freedman W.L., et al. 1994, \apj, 427, 628 		   
\bibitem[xxx]{xxx} Freedman W.L., et al. 2001, \apj, 553, 47
\bibitem[xxx]{xxx} Freeman, K.C., 1996, in ASP Conf. Ser. 92, Formation of 
                   the Galactic Halo: Inside and Out, ed. G.W. Preston, 
		   H. Morrison, \& A. Sarajedini, 3	
\bibitem[xxx]{xxx} Garnavich, P.M., VandenBerg, D.A., Zurek, D.R., 
                    \& Hesser, J.E., 1994, \aj, 107, 1097 
\bibitem[xxx]{xxx} Graham, J.A., 1982, \apj, 252, 474		   		   
\bibitem[xxx]{xxx} Gratton, R.G., Carretta, E., Matteucci, F., \& Sneden, C.,
                   2000, \aap, 358, 671
\bibitem[xxx]{xxx} Gregg, M.D., Ferguson, H.C., Minniti, D., Tanvir, N., 
                   Catchpole, R., 2004, AJ, 127, 1441  
\bibitem[xxx]{xxx} Greggio, L., 1997, \mnras, 285, 151	
\bibitem[xxx]{xxx} Grillmair, C.J., et al. 1996, \aj, 112, 1975  
\bibitem[xxx]{xxx} Guarnieri, M.D., Ortolani, S., Montegriffo, P., Renzini, A.,
                   Barbuy, B., Bica, E., \& Moneti, A., 1998, \aa, 331, 70
\bibitem[xxx]{xxx} Harris, G.H., Harris, W.E., \& Poole, G.B., 1999, \aj,
                   117, 855
\bibitem[xxx]{xxx} Harris, W.E., \& Harris, G.H., 2000, \aj, 120, 2423
\bibitem[xxx]{xxx} Harris, W.E., \& Harris, G.H., 2002, \aj, 123, 3108	
\bibitem[xxx]{xxx} Hartwick, F.D.A., 1976, \apj, 209, 418 
\bibitem[xxx]{xxx} Ho, L.C., Filippenko, A.V., Sargent, W.L.W., 1996, \apj, 
                   462, 183
\bibitem[xxx]{xxx} Holland, S., Fahlman, G.G., \& Richer, H.B., 1996, \aj, 
                   112, 1035	    
\bibitem[xxx]{xxx} Holtzman, J.A., Burrows, C.J., Casertano, S., et al., 
                   1995, PASP, 107, 1065   
\bibitem[xxx]{xxx} Ibata, R.A., Irwin M.J., Lewis, G., Ferguson, A.M.N., 
                   \& Tanvir, N., 2001, Nature, 412, 49
\bibitem[xxx]{xxx} Larsen, S.S., Brodie, J.P., Beasley, M.A., \& Forbes, D.A.,
                   2002, \aj, 124, 828 		
\bibitem[xxx]{xxx} Lee, M.G., Freedman, W.L., \& Madore, B.F. 1993,
                   \apj, 417, 553 
\bibitem[xxx]{xxx} McWilliam, A., 1997, \araa, 35, 503 
\bibitem[xxx]{xxx} Mouhcine, M., Lan\c{c}on, A., 2003, \mnras, 338, 572    	   
\bibitem[xxx]{xxx} Mouhcine, M., Ferguson, H.C., Rich, R.M., Brown, T., 
                   \& Smith E., 2004a, \apj, submitted      	   
\bibitem[xxx]{xxx} Mouhcine, M., Ferguson, H.C., Rich, R.M., Brown, T., 
                   \& Smith E., 2004b, \apj, submitted   
\bibitem[xxx]{xxx} Mould, J., \& Kristian, J., 1986, \apj, 305, 591
\bibitem[xxx]{xxx} Mighell, K.J., Rich, R.M., 1995, \apj, \aj, 110, 1649 
\bibitem[xxx]{xxx} Norris, J.E., \& Ryan, S.G., 1991, \apj, 380, 403 
\bibitem[xxx]{xxx} Origlia, L., Rich, R.M., Castro, S., 2002, \aj, 123, 1559
\bibitem[xxx]{xxx} Pagel, B.E.J., \& Patchett, B.E., 1975, \mnras, 172, 13 
\bibitem[xxx]{xxx} Pagel, B.E.J., 1997, Nucleosyntheis and Chemical Evolution 
                   of Galaxies (Cambridge: Cambridge University Press), 237
\bibitem[xxx]{xxx} Peng, E.W., Holland, C., Freeman, K.C., White, R.L., 2002,
                   \aj, 124, 3144	
\bibitem[xxx]{xxx} Peterson, R.C., Green, E.M., 1998, \apj, 502, 39   		   
\bibitem[xxx]{xxx} Pierce, M.J., \& Tully, R.B., 1992, \apj, 387, 47
\bibitem[xxx]{xxx} Pritchet, C.J., Schade, D., Richer, H.B., Crabtree, D., 
                   \& Yee, H.K.C., 1987, \apj, 323, 79	   
\bibitem[xxx]{xxx} Pritchet, C.J., van den Bergh, S., 1994, \aj, 107, 1730 
\bibitem[xxx]{xxx} Puche, A., \& Carignan, C., 1988, \aj, 95, 1025
\bibitem[xxx]{xxx} Richer, H.B., 1981, \apj, 243, 744
\bibitem[xxx]{xxx} Richer, H.B., Pritchet, C.J., Crabtree, D.R., 1985, \apj, 
                   298, 240
\bibitem[xxx]{xxx} Robin, A.C., Reyl\'e, C., Derri\'ere, S., \& Picaud, S.,
                   2003, \aap, 409, 523	   
\bibitem[xxx]{xxx} Sagar, R., Subramaniam, A., Richtler, T., \& Grebel, E.K. 
                   1999, \aaps, 135, 391
\bibitem[xxx]{xxx} Sakai, S., Mould J.R., Hughes S.M.G., et al., 2000, \apj, 
                   529, 698  	   
\bibitem[xxx]{xxx} Sarajedini, A., \& Van Duyne, J., 2001, \aj, 122, 2444   
\bibitem[xxx]{xxx} Saviane, I., Rosenberg, A., Piotto, G., \& Aparicio, A., 
                   2000, \aa, 355, 966 
\bibitem[xxx]{xxx} Schlegel, D.J., Finkbeiner, D.P., \& Davis, M., 1998,
                   \apj, 500, 525
\bibitem[xxx]{xxx} Scott, D. W. 1992, Multivariate Density Estimation 
                   (New York: Wiley)
\bibitem[xxx]{xxx} Searle, L., \& Sargent, W.L.W., 1972, \apj, 173, 25 
\bibitem[xxx]{xxx} Searle, L., \& Zinn R., 1978, \apj, 225, 357  
\bibitem[xxx]{xxx} Shetrone, M., Venn, K.A., Tolstoy, E., Primas, F., Hill, V.,
                   \& Kaufer., A., 2003, \aj, 125, 684  
\bibitem[xxx]{xxx} Smail, I., Hogg, D.W., Yan, L., \& Cohen J.G., 1995, \apj,
                   449, L105
\bibitem[xxx]{xxx} Taylor, B.J., 2001, \aap, 377, 473	   
\bibitem[xxx]{xxx} Tiede, G.P., Sarajedini, A., Barker, M.K.,  2004, 
                   \aj, 128, 224
\bibitem[xxx]{xxx} Tinsley B.M., 1980, Fundam. Cosmic Phys., 5, 287
\bibitem[xxx]{xxx} Tolstoy, E., Venn, K.A., Shetrone, M., Primas, F., Hill, V.,
                   \& Szeifert, T., 2003, \aj, 125, 707
\bibitem[xxx]{xxx} VandenBerg, D.A., Swenson, F.J., Rogers, F.J., 
                   Iglesias, C.A., \& Alexander, D.R,. 2000, \apj, 532, 430
\bibitem[xxx]{xxx} Yanny, B., Newberg, H.J., Kent, S., et al., 2000, \apj, 
                   540, 825
\bibitem[xxx]{xxx} Zoccali, M., Renzini, A., Ortolani, S., et al., 2001, \aj, 
                   121, 2638    
\bibitem[xxx]{xxx} Zoccali, M., Renzini, A., Ortolani, S., et al., 2003, \aap, 
                   399, 931
\end{thebibliography}
\end{document}